%% file: paper_v16.tex
\documentclass{aa}  
\usepackage{natbib,epsfig,graphicx,color,psfrag,amsmath,pdflscape}
\usepackage[varg]{txfonts}
\bibpunct{(}{)}{,}{a}{}{,}
\usepackage{multicol}
\usepackage{multirow}
\include{definitions}

\begin{document}
\title{Binary-object spectral-synthesis in 3D (\textsc{BOSS-3D})}
\subtitle{Modelling \Ha~emission in the enigmatic multiple system LB-1}

\author{L. Hennicker\inst{\ref{leuven}} \and N. D. Kee\inst{\ref{hawaii}} \and T. Shenar\inst{\ref{leuven}, \ref{amsterdam}} \and J. Bodensteiner\inst{\ref{leuven}, \ref{munich}} \and M. Abdul-Masih\inst{\ref{chile}} \and I. El Mellah\inst{\ref{france}} \and H. Sana\inst{\ref{leuven}} \and J. O. Sundqvist\inst{\ref{leuven}}}

\institute{Institute of Astronomy, KU Leuven, Celestijnenlaan 200D,
  3001 Leuven, Belgium
  \\ \email{levin.hennicker@kuleuven.be}\label{leuven}
  \and
   National Solar Observatory, 22 Ohi'a Ku Street, Makawao, HI 96768, USA\label{hawaii}  
   \and
   Anton Pannekoek Institute for Astronomy, University of Amsterdam, Postbus 94249, 1090 GE Amsterdam, The Netherlands\label{amsterdam}
   \and
   European Southern Observatory, Karl-Schwarzschild-Strasse 2, D-85748 Garching bei M\"unchen, Germany\label{munich}
  \and      
  European Southern Observatory, Alonso de Cordova 3107, Vitacura, Casilla 19001, Santiago de Chile, Chile\label{chile}
  \and      
  Institut de Plan\'{e}tologie et d'Astrophysique de Grenoble, 414 Rue de la Piscine, 38400 Saint-Martin-d'H\`{e}res, France\label{france}
}
\date{Received 20 July 2021; Accepted 23 November 2021}
\abstract 
{To quantitatively decode the information stored within an observed
  spectrum, detailed modelling of the physical state and accurate
  radiative transfer solution schemes are required. The accuracy of
  the model is then typically evaluated by comparing the calculated
  and observed spectra. In the analysis of stellar spectra, the
  numerical model often needs to account for binary companions and 3D
  structures in the stellar envelopes. The enigmatic binary (or
  multiple) system \mbox{LB-1} constitutes a perfect example of such a
  complex multi-D problem. Thus far, the \mbox{LB-1} system has
  been indirectly investigated by 1D stellar-atmosphere
  codes and by spectral disentangling techniques, yielding differing
  conclusions about the nature of the system (\eg~a B-star and black-hole
  binary with an accretion disc around the black hole or a
  stripped-star and Be-star binary system have been proposed).}
{To improve our understanding of the \mbox{LB-1} system, we directly modelled
  the phase-dependent \Ha~line profiles of this system. To this end,
  we developed and present a multi-purpose binary-object
  spectral-synthesis code in 3D (\textsc{BOSS-3D}).}
{\textsc{BOSS-3D} calculates synthetic line profiles for a given state
  of the circumstellar material. The standard $pz$-geometry commonly
  used for single stars is extended by defining individual coordinate
  systems for each involved object and by accounting for the appropriate
  coordinate transformations. The code is then applied to the \mbox{LB-1}
  system, considering two main hypotheses, a binary containing a
  stripped star and Be star, or a B star and a black hole with a
  disc.}
{Comparing these two scenarios, neither model can reproduce the
  detailed phase-dependent shape of the \Ha~line profiles. A
  satisfactory match with the observations, however, is obtained by
  invoking a disc around the primary object in addition to the Be-star
  disc or the black-hole accretion disc.}
{The developed code can be used to model synthetic line profiles for a
  wide variety of binary systems, ranging from transit spectra of
  planetary atmospheres, to post-asymptotic giant branch binaries
  including circumstellar and circumbinary discs and massive-star
  binaries with stellar winds and disc systems. For the \mbox{LB-1} system,
  our modelling provides strong evidence that each object in the
  system contains a disc-like structure.}
\titlerunning{Binary-object spectral-synthesis in 3D}
\authorrunning{L. Hennicker et al.}
\maketitle
%
%
\section{Introduction} 
\label{sec:intro}
The interpretation of observed radiation emerging, for instance, from
galaxies, individual stars, or planetary systems, is key to
understanding fundamental properties of the Universe. Deep insight
into the underlying physics of an observed object can be obtained by
analysing the electromagnetic spectrum of the emitting
object. Decoding the information stored within a spectrum, however, is
not a simple task. Deducing the fundamental parameters of a single
star (\ie~\Lstar, \Teff, \Rstar), for example, requires detailed
modelling of the stellar atmosphere, possibly relaxing the assumption
of local thermal equilibrium (LTE) and accounting for wind
outflows. For a given model atmosphere, synthetic spectra then need to
be calculated and compared with the observations. In the OB-star
regime (on which this paper focusses), state-of-the-art atmospheric
modelling and spectral synthesis codes typically assume spherical
symmetry (\eg~{\sc PHOENIX}: \citealt{Haus92}; {\sc CMFGEN}:
\citealt{hilliermiller98}, \citealt{Hillier2012}; WM-{\sc basic}:
\citealt{pauldrach01}; {\sc PoWR}: \citealt{Hamann2003},
\citealt{Sander2017b}; {\sc FASTWIND}: \citealt{Sundqvist2018b},
\citealt{Puls2020}).

Many stars, however, can deviate from spherical symmetry, with
perturbations typically induced by magnetic fields, surface (and wind)
distortions from rotation, surrounding discs, and
binarity or multiplicity effects. Due to the complexity of including full
non-LTE occupation numbers within multi-D calculations, corresponding
spectral synthesis codes are only gradually developed by applying
different solution methods from finite-volume methods (FVM) to short-
and long-characteristics methods (SC and LC) to Monte-Carlo methods
(MC). A non-exhaustive list of 3D radiative transfer codes accounting
for supersonic velocity fields includes \textsc{Wind3D}
(\citealt{Lobel08}, FVM), a code developed by \citealt{Hennicker2020}
(SC), \textsc{Phoenix/3D} (\citealt{Haus06} and other papers in this
series, LC), and \textsc{HDUST} (\citealt{Carciofi2006}, MC), where the
first two examples currently apply a two-level-atom (TLA) approach,
whereas the last two include a multi-level description of an atom (of
typically one particular species). When the atomic level populations
are obtained by such codes, the synthetic spectra can be calculated by
solving the equation of radiative transfer along the direction to the
observer, for instance, by applying the LC method in a cylindrical
coordinate system based on \cite{Lamers1987}, \cite{Busche2005},
\cite{Sundqvist12c} (see also \citealt{Hennicker2018}).

Thus far, the above described codes and methods have been tailored to the
spectral synthesis of single stars in 1D or 3D. To our knowledge,
there exists no viable multi-D alternative to calculating synthetic
spectra from massive-star binary systems with circumstellar material
around them. For example, the \textsc{SPAMMS}-code by
\cite{AbdulMasih2020b} relies on patching the emergent intensities as
obtained from several 1D spherically symmetric \textsc{FASTWIND}
models. Thus, the effect of rays propagating through the winds of both
individual stars is neglected. Moreover, such models cannot be used
for arbitrary 3D structures such as circumstellar discs.

In this paper, we thus develop a general purpose spectral synthesis
tool for binary systems (\textsc{BOSS-3D}, \ie~binary-object
spectral-synthesis in 3D) by extending the solution method for single
stars from \cite{Hennicker2018} using simple coordinate
transformations. For given atmospheric and circumstellar structure(s)
with (thus far approximate) occupation numbers, this code accounts for
large-scale, possibly structured outflows or discs within both
involved systems, thus circumventing the aforementioned shortcomings
of previous methods.  By allowing for different length scales of the
involved objects, the code can also be used to analyse transit spectra
of planetary systems.  The basic idea of our method relies on defining
two (independent) coordinate systems for the individual objects, and
applying the standard single-star $pz$-geometry to each of the
systems. To correctly account for overlapping coordinates (\eg~during
transits), the individual systems are merged by triangulation.

As a first application of the newly developed algorithm, we considered
the binary (or multiple) system \mbox{LB-1}. The spectrum of this
system clearly shows an anti-phase behaviour of stellar absorption
lines against the line wings of a broad \Ha~emission
(\citealt{Liu2019}). Yet the interpretation of these findings is still
under debate. Based on these observations and the corresponding
orbital reconstruction, \cite{Liu2019} proposed a binary system with
period $P\approx 79\,{\rm d}$, consisting of a B star (visible
component) and a $70\,\msun$ black hole (BH) with associated disc
yielding the \Ha~emission.  Since it is difficult to explain the
origin of a $70\,\msun$ BH in the Milky Way from stellar evolution
theory (though, see \citealt{Belczynski2020}), there has been vivid
discussion about \mbox{LB-1} in the literature. While
\cite{SimonDiaz2020} re-analysed the visible component (B star,
primary object) and found a significantly lower mass for the B star
thus also reducing the BH mass, \cite{AbdulMasih2020} and
\cite{ElBadry2020} showed that the wobbling \Ha~wings could also be
explained by a superposed broad \Ha~absorption component on a static
emission profile, thus questioning the BH hypothesis as a
whole. \cite{Shenar2020} disentangled the optical spectra of
\mbox{LB-1} and proposed a binary system consisting of a stripped
He-star (the primary) in a $79\,{\rm d}$ orbit with a rapidly-rotating
Be-star, presumably the product of a recent mass-transfer
event. Alternatively, \cite{Rivinius2020} proposed that the Be-star is
a tertiary object, leaving the secondary component
unidentified. Recently, \cite{Lennon2021} modelled the spectral energy
distribution (SED) of \mbox{LB-1} in the optical and ultra-violet
regime using 1D plane-parallel model-atmospheres, and compared several
models (including the aforementioned stripped-star/Be-star and
B-star/BH hypotheses) with observations from the Hubble Space
Telescope. While their models slightly favour the B-star and BH scenario,
the stripped-star and Be-star hypothesis could not be ruled out entirely.
Throughout this paper, these two competing scenarios will be
abbreviated by the B+BH and the strB+Be scenario, respectively.

In this paper, we aim to provide further constraints on \mbox{LB-1} by
applying the \textsc{BOSS-3D} code to models that represent the main
hypotheses detailed above, namely the B+BH and the strB+Be binary
scenarios. By calculating the corresponding \Ha~line profiles, we show
that both hypotheses can explain the observed phase-dependent \Ha~line
profile provided the \mbox{LB-1} system contains an additional disc
attached to the B-star primary (\ie~either to the B star in the B+BH
scenario, or to the stripped star in the strB+Be scenario). The paper
is structured as follows. In Sect.~\ref{sec:single_theory} we review
the basic numerical techniques for spectral synthesis of single
objects, in Sect.~\ref{sec:binary_theory} we extend these methods to
binary systems, and in Sect.~\ref{sec:lb1} we apply the developed code
to the \mbox{LB-1} system. Finally, we summarize our findings in
Sect.~\ref{sec:conclusions}.
%
%
\section{Single-star spectral synthesis}
\label{sec:single_theory}
\begin{figure*}[t]
  \centering
  \resizebox{0.33\hsize}{!}{\includegraphics{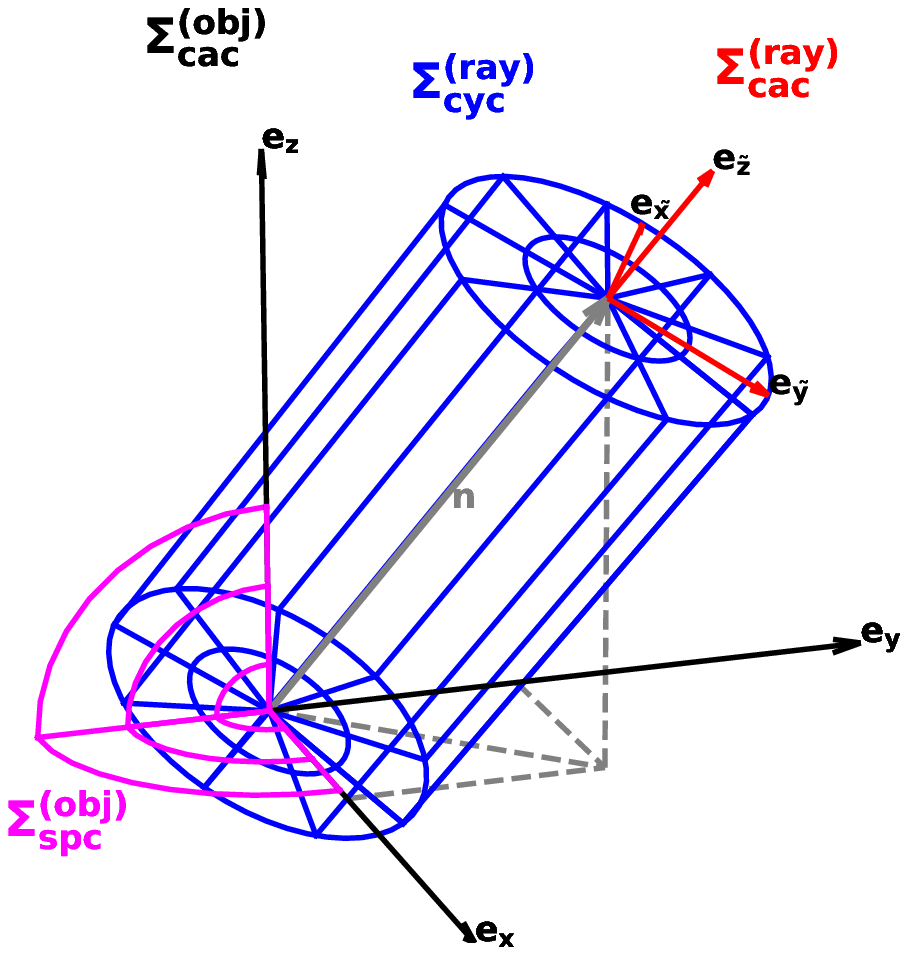}}
  \resizebox{0.32\hsize}{!}{\includegraphics{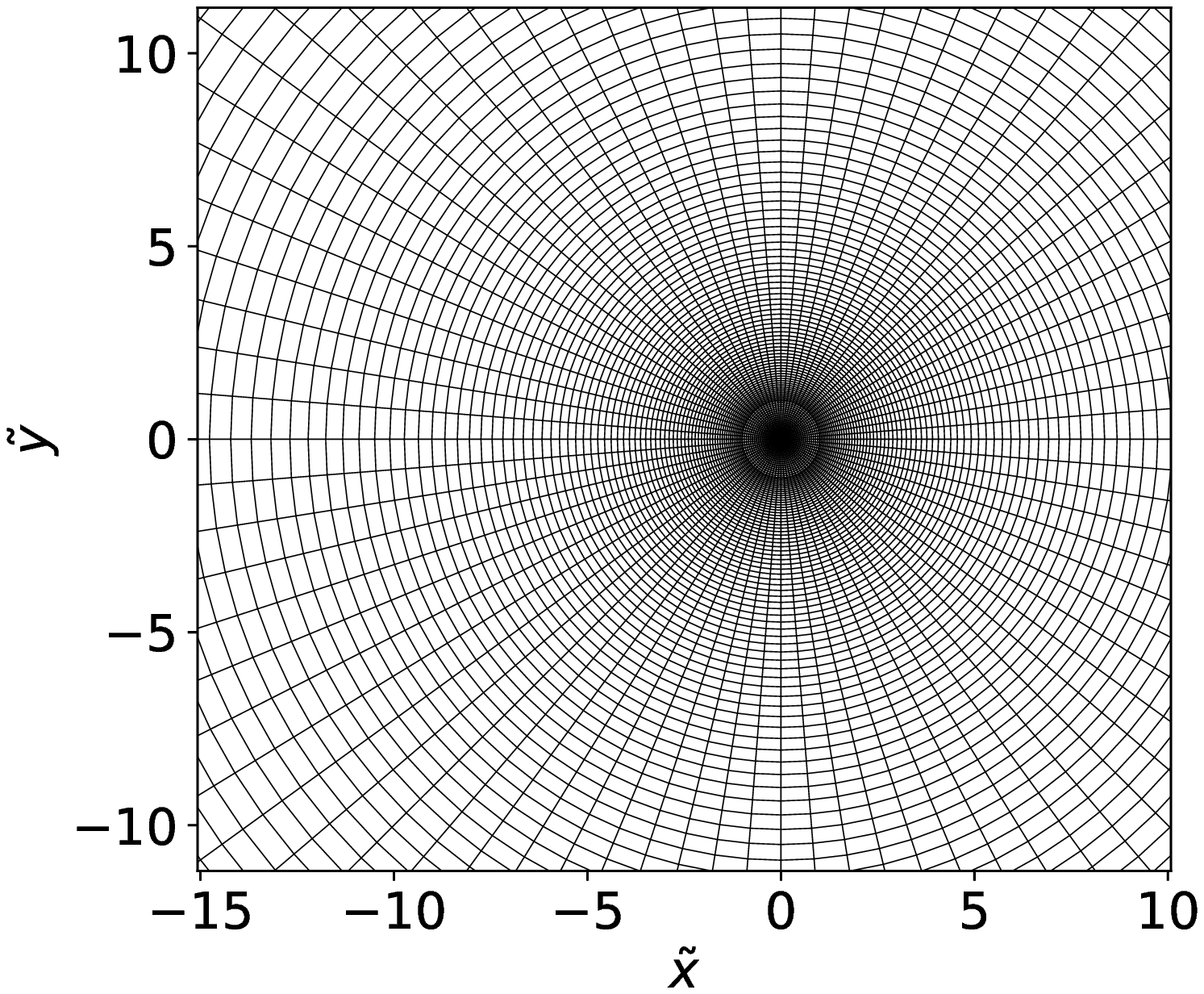}}  
  \resizebox{0.34\hsize}{!}{\includegraphics{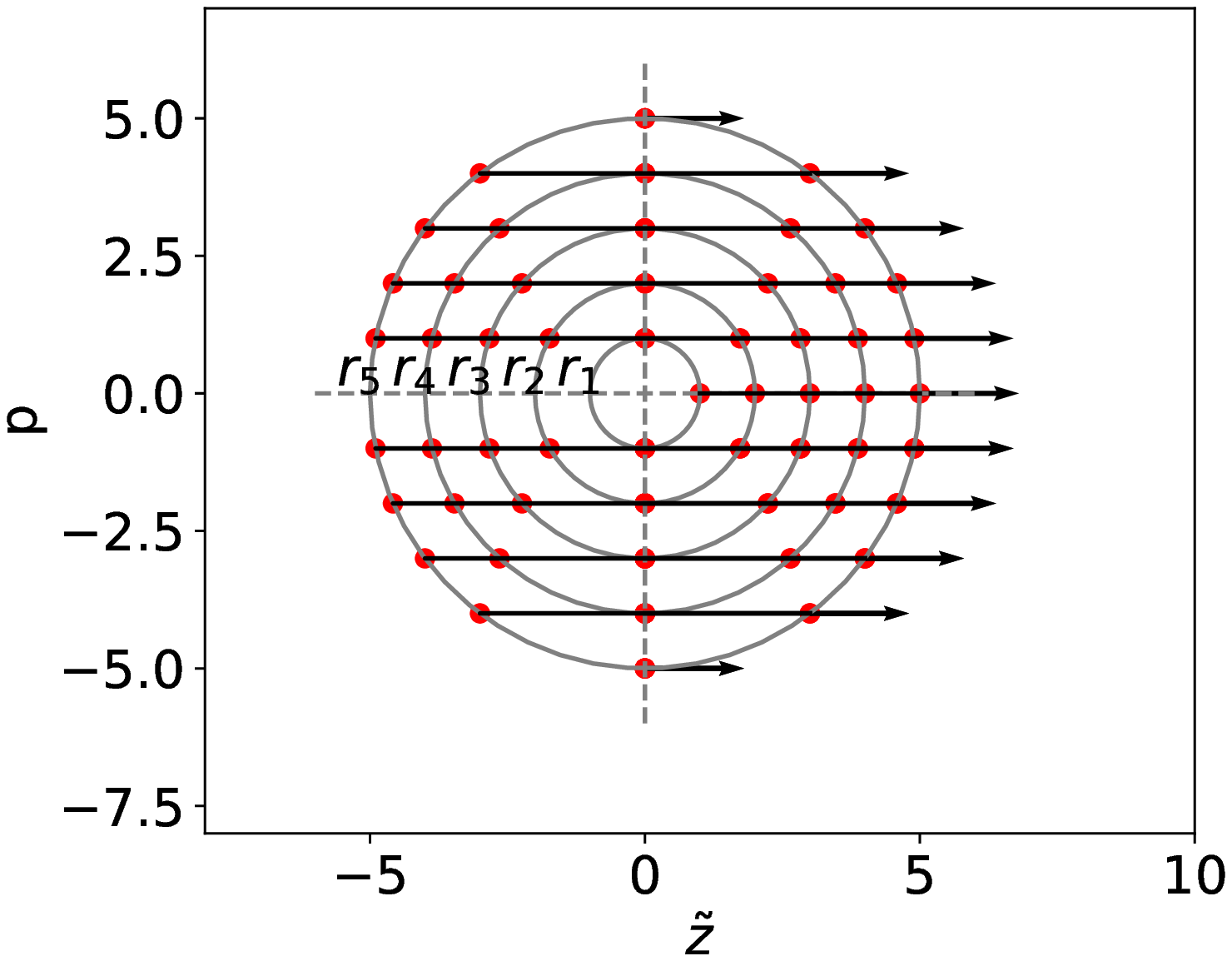}}
  \resizebox{0.33\hsize}{!}{\includegraphics{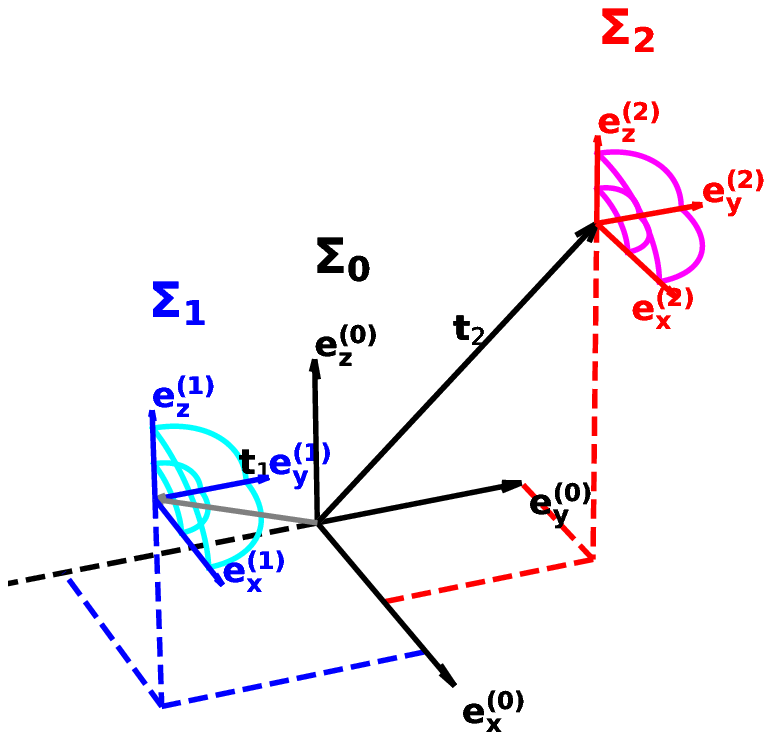}}
  \resizebox{0.32\hsize}{!}{\includegraphics{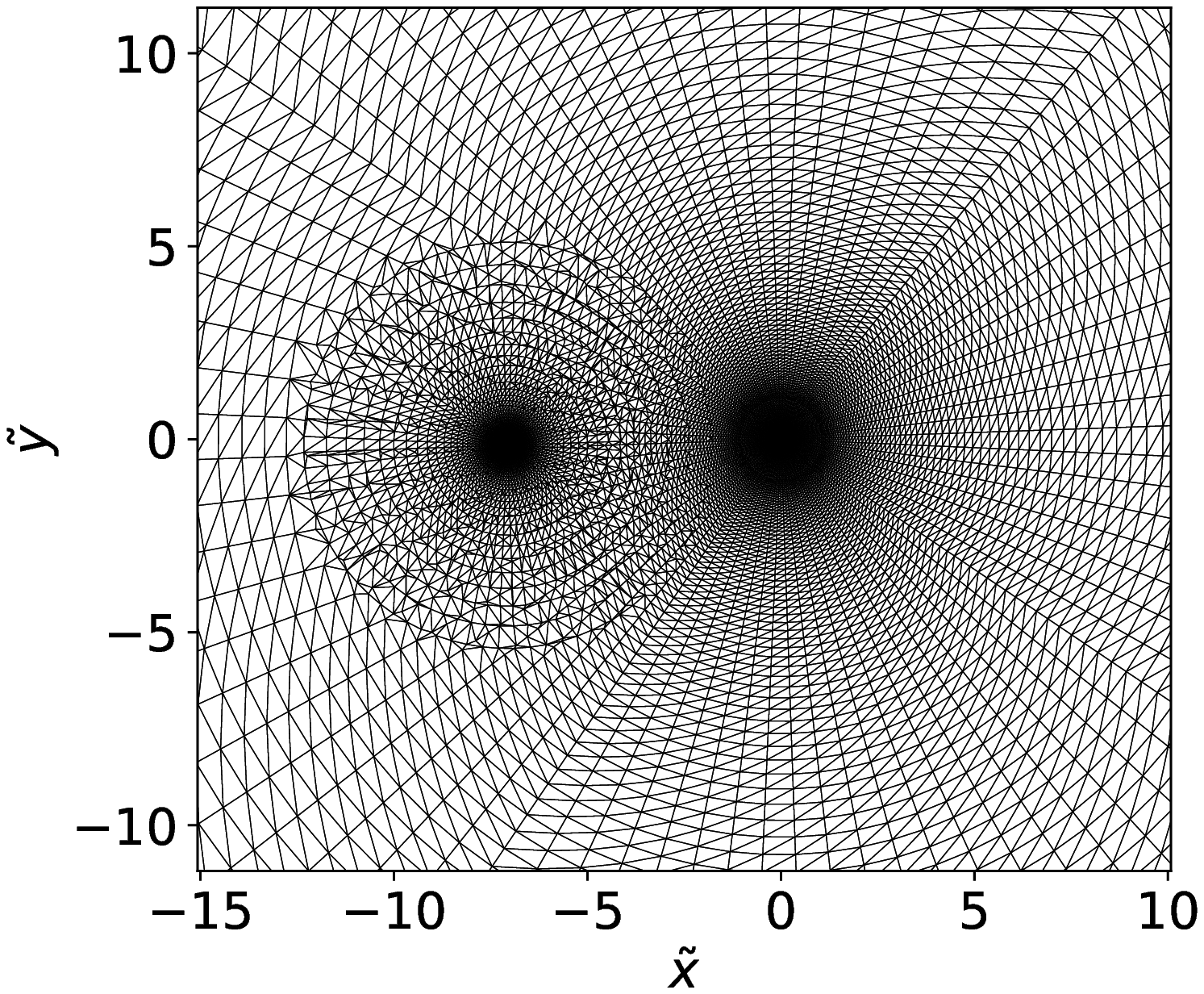}}
  \resizebox{0.34\hsize}{!}{\includegraphics{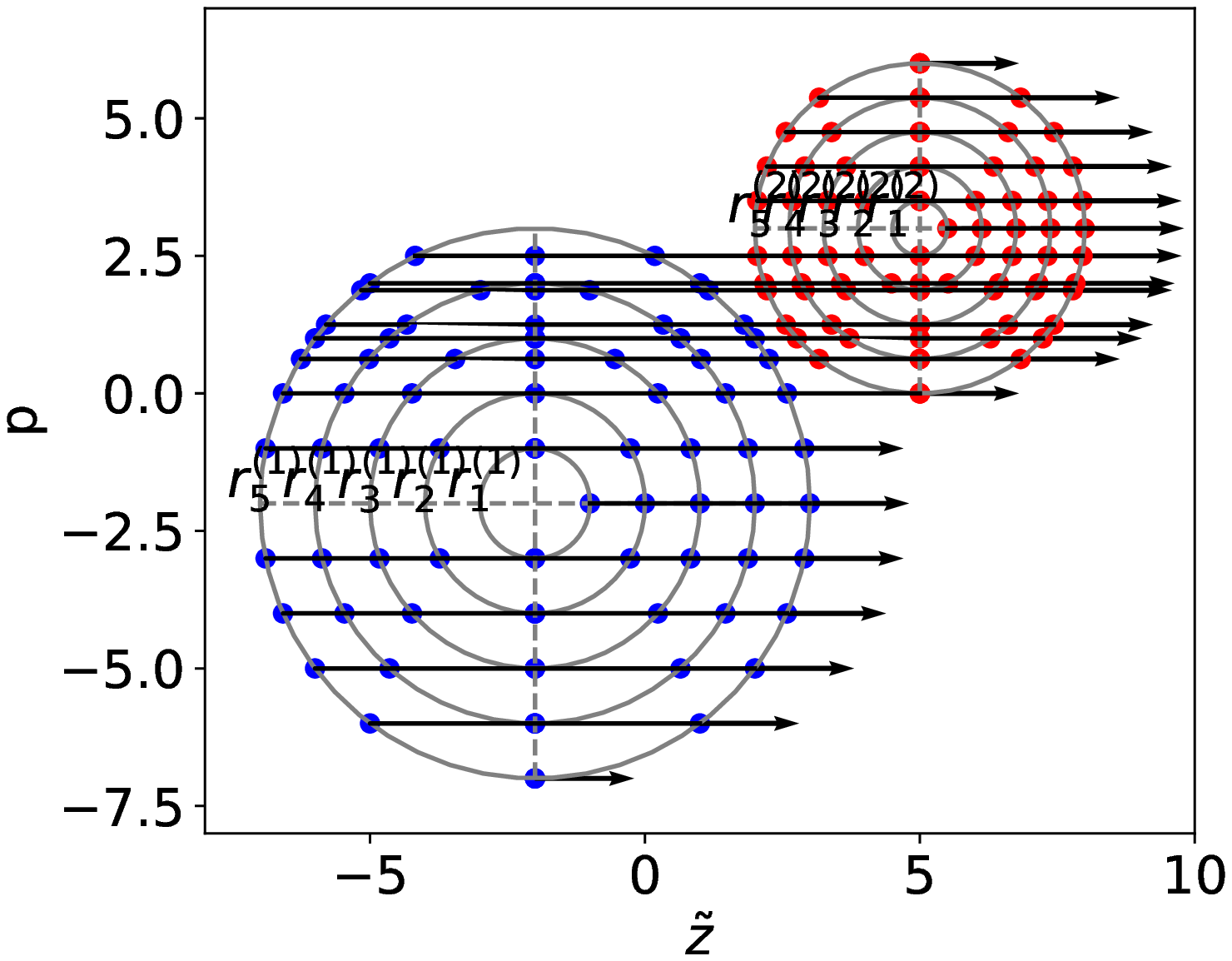}}   
 \caption{Geometry used for the single-star algorithm (top panels) and
   for the binary code (bottom panels). Top left panel: Cylindrical
   coordinate system $\Sigma_{\rm cyc}^{(\rm{ray})}$ (blue) with
   corresponding Cartesian reference frame $\Sigma_{\rm
     cac}^{(\rm{ray})}$ (red) for an observer's direction
   $\vecown{n}\parallel \vecown{e}_{\tilde{z}}$ used to obtain the
   synthetic spectrum of a given (single) object described in a
   spherical coordinate system $\Sigma_{\rm spc}^{(\rm{obj})}$
   (magenta) with corresponding Cartesian reference frame $\Sigma_{\rm
     cac}^{(\rm{obj})}$ (black).  Top middle panel: The $p-\zeta$ (or
   $\tilde{x}-\tilde{y}$-plane) of the cylindrical coordinate system
   (perpendicular to the observer's direction).  Top right panel: An
   arbitrary $p-\tilde{z}$-slice, with $\vecown{e}_{\tilde{z}}$
   pointing towards the observer. The grey circles indicate the radial
   grid of the spherical coordinate system of a considered (single)
   object, and the black arrows correspond to the individual rays for
   each impact parameter $p$. The radiative transfer is then performed
   along each ray with corresponding discretized
   $\tilde{z}_{\indx{k}}$ coordinates indicated by the red
   dots. Bottom left panel: Two local coordinate systems $\Sigma_1$
   and $\Sigma_2$ describing two individual objects indicated by the
   cyan and magenta quarter-circles, respectively. Both coordinate
   systems are embedded within a global coordinate system $\Sigma_0$
   (typically but not necessarily chosen to be the centre of mass of
   both objects). The vectors $\vecown{t}_1$ and $\vecown{t}_2$ describe the
   origin of the local coordinate systems in the global coordinate
   system.  Bottom middle panel: As top middle panel, however showing
   the triangulation of the $\tilde{x}-\tilde{y}$-plane for an
   arbitrary orbital configuration in a global cylindrical coordinate
   system within the binary algorithm.  Bottom right panel: As top
   right panel, but now for the binary system. The individual
   $\tilde{z}$-rays propagating through the system of object 1 (blue)
   and object 2 (red) potentially need to be merged.}
%
\label{fig:transform01}
\end{figure*}
In this section, we briefly summarize the basic solution method for
obtaining synthetic spectra from a given single-star model in 3D.
Following, for instance, \cite{Busche2005}, the radiation flux at
frequency $\nu$ emerging from any unresolved emitting object as seen
by an observer at distance $d$ can be formulated as
\beq
\label{eq:fluxint}
F_\nu = \dfrac{1}{d^2} \int_0^{2\pi} \int_0^{R_{\rm max}} \Inu \left(p,\zeta,\tilde{z}=R_{\rm max} \right) p \dd p \dd \zeta \,,
\eeq
where $R_{\rm max}$ is the maximum size of the emitting object,
$\left( p, \zeta\right)$ are polar coordinates describing the
projected disc of the emitting object perpendicular to the observer's
direction $\vecown{n}$, and $I_\nu$ is the emergent specific intensity
into that direction evaluated at a distance $\tilde{z}=R_{\rm max}$.

The flux can then be obtained by numerically integrating the emergent
specific intensity over the projected disc, with the intensity to be
calculated by solving the equation of radiative transfer along all
rays in the discretized $(p_i,\zeta_j )$ domain of a cylindrical
coordinate system $\Sigma_{\rm cyc}^{({\rm ray})} =
\left(p,\zeta,\tilde{z} \right)$ with corresponding Cartesian
reference frame $\Sigma_{\rm cac}^{({\rm ray})} =
\left(\tilde{x},\tilde{y},\tilde{z} \right)$, and with the
$\tilde{z}$-axis being aligned with the direction to the observer (see
also Fig.~\ref{fig:transform01}).  The corresponding
(time-independent) equation of radiative transfer reads:
\beq
\label{eq:eqrt}
\dfrac{\dd I_\nu\left(p_i,\zeta_j ,\tilde{z}\right)}{\dd \tilde{z}} = \chi_\nu \left(p_i,\zeta_j,\tilde{z}\right) \left[
S_\nu\left(p_i,\zeta_j,\tilde{z} \right) - I_\nu\left(p_i,\zeta_j,\tilde{z} \right) \right] \,,
\eeq
with the opacity $\chi_\nu$ and source function $S_\nu$ describing the
absorption and emission of all continuum and line processes. For
simplicity, we neglect continuum processes in the following and only
consider a single line transition between lower and upper level
$l\leftrightarrow u$ with transition frequency $\nu_{lu}$.  Further,
we omit the notation for explicit spatial dependencies.  Then, the
opacity and source functions are given by:
\beqa
\label{eq:defopal}
\chi_\nu &=& \dfrac{\pi \echarge^2}{\mel c}\left(gf\right) \left[\dfrac{n_l}{g_l}-\dfrac{n_u}{g_u} \right] \Phi_{lu} \left(\nu \right) \\
\label{eq:defsline}
S_\nu &=& \dfrac{2 h \nu_{lu}^3}{c^2} \dfrac{1}{\dfrac{n_l}{n_u}\dfrac{g_l}{g_u}-1} \,,
\eeqa
where $n_l$ and $n_u$ are the occupation numbers, $g_l$ and $g_u$ are
the statistical weights, and
\beq
\Phi_{lu}\left(\nu \right) = \dfrac{1}{\sqrt{\pi}\ddop} \exp \left[-\left(\dfrac{\nu - \nu_{lu}}{\ddop} - \dfrac{\vecown{n}\cdot\vecown{v}}{\vth}\right)^2 \right]
\eeq
is the profile function (approximated here by a Doppler profile) at
observers frame frequency $\nu$ including non-relativistic
  Doppler shifts from the observer's to the comoving frame via the
  projected velocity $\vecown{n}\cdot\vecown{v}$. The width of the
profile, $\ddop = \nu_{lu} \dfrac{\vth}{c}$, is given from the thermal
velocity of the considered species with mass number $m_{\rm A}$,
accounting for the local temperature $T$ and a micro-turbulent
velocity $\vmicro$:
\beq
\vth = \sqrt{\dfrac{2 \kboltz T}{m_{\rm A}} + \vmicro^2}\,.
\eeq
For a given atmospheric model (with known temperatures, velocity
vectors, occupation numbers, and micro-turbulent velocities), the
equation of radiative transfer, Eq.~\eqref{eq:eqrt}, becomes an
ordinary differential equation with exact solution from any point
$\tilde{z}_{k-1}$ to $\tilde{z}_k$
\beq
\label{eq:formal_solution}
I_\nu \left(\tilde{z}_{k}\right) = I_\nu \left(\tilde{z}_{k-1} \right) \eu^{-\Delta \tau_{k}} + \eu^{-\Delta \tau_{k}} \int_0^{\Delta \tau_{k}} e^t S(t) \dd t \,.
\eeq
We note that $\Delta \tau_k := \int_{\tilde{z}_{k-1}}^{\tilde{z}_{k}} \chi_\nu \dd
\tilde{z}$ is the optical-depth increment between two positions along
a particular ray. Equation \eqref{eq:formal_solution} is discretized by
approximating the source function by a constant, linear, or quadratic
form in optical-depth space (\eg~\citealt{Hennicker2020}, Eq. 12)
based on \citet[][linear and quadratic interpolations]{Kunasz88} and
\citet[][monotonic B\'ezier interpolations]{Hayek2010}.

Since we expect the input model to be given in spherical coordinates
$\Sigma_{\rm spc}^{({\rm obj})}=(r,\clatitude,\azimuth)$ with
corresponding Cartesian reference frame $\Sigma_{\rm cac}^{({\rm
    obj})}=(x,y,z)$, and if we assume that all structures are well
resolved by the corresponding discretized grid, the discretization of
$\tilde{z}_k$ is easily constructed following the standard
$pz$-geometry approach (\eg~\citealt{mihalasbook78}, see also
Fig.~\ref{fig:transform01}):
\beq
\label{eq:zgrid}
\tilde{z}_k \in
\begin{cases}
\left[-\sqrt{r_q^2-p_i^2}, \sqrt{r_q^2-p_i^2} \right] \quad \forall r_q \geq \lvert p_i \rvert \quad {\rm and} \quad r_q \geq \Rstar,\\
\left[\sqrt{\Rstar^2-p_i^2}, \sqrt{r_q^2-p_i^2} \right] \,\,\quad \forall r_q \geq \lvert p_i \rvert \quad {\rm and} \quad r_q < \Rstar
\end{cases}
\,,
\eeq
where $r_q$ refers to the discretized grid of the spherical coordinate
system.  Thus, the required coordinates for each ray corresponding to
a certain $(p_{\indx{i}},\zeta_{\indx{j}})$ pair are determined in the
cylindrical coordinate system $\Sigma_{\rm cyc}^{(\rm{ray})}$. All
quantities describing the state of the gas along a ray need to be
interpolated from the spherical grid of the input model. To this end,
we need to transform coordinates from $\Sigma_{\rm
  cyc}^{(\rm{ray})}\rightarrow \Sigma_{\rm cac}^{(\rm{ray})}
\rightarrow \Sigma_{\rm cac}^{(\rm{obj})} \rightarrow \Sigma_{\rm
  spc}^{(\rm{obj})}$:
\beqa
\label{eq:transforms01}
\begin{pmatrix}
  \tilde{x} \\
  \tilde{y} \\
  \tilde{z} 
\end{pmatrix}
=
\begin{pmatrix}
  p \cos \zeta \\
  p \sin \zeta \\
  \tilde{z} 
\end{pmatrix}
\\
\label{eq:transforms02}
\begin{pmatrix}
  x \\
  y \\
  z 
\end{pmatrix}
=
\begin{pmatrix}
\vecown{e}_{\tilde{x}} \cdot \vecown{e}_x & \vecown{e}_{\tilde{y}} \cdot \vecown{e}_x & \vecown{e}_{\tilde{z}} \cdot \vecown{e}_x \\
\vecown{e}_{\tilde{x}} \cdot \vecown{e}_y & \vecown{e}_{\tilde{y}} \cdot \vecown{e}_y & \vecown{e}_{\tilde{z}} \cdot \vecown{e}_y \\
\vecown{e}_{\tilde{x}} \cdot \vecown{e}_z & \vecown{e}_{\tilde{y}} \cdot \vecown{e}_z & \vecown{e}_{\tilde{z}} \cdot \vecown{e}_z
\end{pmatrix}
\cdot
\begin{pmatrix}
  \tilde{x} \\
  \tilde{y} \\
  \tilde{z} 
\end{pmatrix}
=\matown{A}\cdot \tilde{\vecown{r}}
\\
\label{eq:transforms03}
\begin{pmatrix}
  r \\
  \clatitude \\
  \azimuth 
\end{pmatrix}
=
\begin{pmatrix}
  \sqrt{x^2+y^2+z^2} \\
  \tan^{-1} \left(\sqrt{x^2+y^2}/z \right) \\
  \tan^{-1} \left(y/x \right)
\end{pmatrix}
\,,
\eeqa
with $\vecown{e}_{x,y,z}$ and
$\vecown{e}_{\tilde{x},\tilde{y},\tilde{z}}$ the unit vectors of the
Cartesian reference frames for the spherical and cylindrical systems,
respectively, and $\matown{A}$ the transformation matrix from the
Cartesian ray system to the object's Cartesian reference frame. The
unit vectors $\vecown{e}_{\tilde{x}}$ and $\vecown{e}_{\tilde{y}}$ can
be chosen arbitrarily under the constraint that
$\vecown{e}_{\tilde{z}} \parallel \vecown{n}$ and
$\vecown{e}_{\tilde{x}} \perp \vecown{e}_{\tilde{y}} \perp
\vecown{e}_{\tilde{z}}$. With the obtained coordinates in the
spherical system of the object, all quantities on the discretized ray
are calculated by tri-linear interpolations. To ensure that the
line-profile function along a ray is resolved by the $\tilde{z}$-grid,
we refine the $\tilde{z}$-grid if the projected velocity steps are
above $\vth/3$.

For a single object, synthetic line profiles are thus obtained for a
specific observer's direction by performing the following steps:
\begin{enumerate}
\item Definition of the $(p,\zeta)$ grid, and of the transformation
  matrix relating the Cartesian reference frames of the cylindrical
  (ray) and spherical (object) coordinate systems.
\item Calculation of the $\tilde{z}_{\indx{k}}$-grid for each
  $(p_{\indx{i}},\zeta_{\indx{j}})$ following Eq.~\eqref{eq:zgrid}.
\item Transformation of the coordinates
  $(p_{\indx{i}},\zeta_{\indx{j}},\tilde{z}_{\indx{k}})$ to the
  object's spherical system by
  Eqs.~\eqref{eq:transforms01}-\eqref{eq:transforms03}, and
  interpolation of all required quantities (\ie~opacities, source
  functions, and velocity components).
\item Solving the discretized form of the radiative transfer equation,
  Eq.~\eqref{eq:formal_solution}, along each ray to obtain the
  emerging specific intensity. Boundary conditions are specified by
  photospheric line profiles for the
  specific intensity if a ray hits the stellar core
  (implemented via a user-specified subroutine to couple
    either line profiles as obtained from Kurucz atmosphere models,
    \eg~\citealt{Castelli2003}, or from
    \textsc{FASTWIND})\footnote{Alternatively, one could
      also explicitly solve the radiative transfer through a
      pre-specified density and temperature stratification in the
      photosphere, of course.}. For non-core rays, the incident
  intensity at the outer boundary is set to zero.
\item Integration of emerging specific intensities over the
  $(p,\zeta)$-plane by replacing the integral in
  Eq.~\eqref{eq:fluxint} with a discretized sum, in order to obtain
  the flux at a specific frequency bin.
\item Restarting the procedure at step 2 for the next frequency bin.
\end{enumerate}
%
%
\section{Multi-object spectral synthesis} 
\label{sec:binary_theory}
In this section, we extend the solution scheme for a single object
described above to multiple systems by accounting for the appropriate
coordinate transformations and triangulation techniques.
When accounting for a secondary object, the algorithm above
essentially suffers from one key issue, namely the resolution of both
involved objects. For instance, the spherical grid of the primary
object would not resolve the secondary object in the required detail
when simply adding the secondary into the spherical grid. Although a
description in Cartesian coordinates (presumably requiring
mesh-refinement techniques) might help here, we follow a different
path within this paper.

Our method is based on the simple observation that each individual
object in a multiple system can be described on an individual
spherical grid tailored to the properties (\eg~length scale) of the
object (see Fig.~\ref{fig:transform01}). All spherical grids are then
embedded within a global coordinate system with origin chosen at the
centre of mass of the multiple system. To keep track of the coordinate
representations for each object $q$, we define the following
transformations between the global system $\Sigma_{{\rm
    cac},0}^{(\rm{obj})}$ and the individual systems $\Sigma_{{\rm
    cac},q}^{(\rm{obj})}$ (see Appendix~\ref{app:transformations}):
\beqa
\label{eq:transform01}
\vecown{r}_{{\rm cac},0}^{(\rm{obj})} &=& \dfrac{L_q}{L_0} \matown{Q}_q
\cdot \vecown{r}_{{\rm cac},q}^{(\rm{obj})} + \vecown{t}_q 
\\
\label{eq:transform02}
\vecown{r}_{{\rm cac},q}^{(\rm{obj})} &=& \dfrac{L_0}{L_q} \matown{Q}_q^{-1}
\cdot \left(\vecown{r}_{{\rm cac},0}^{(\rm{obj})} - \vecown{t}_q \right) \,,
\eeqa
where $\vecown{r}_{{\rm cac},0}^{(\rm{obj})}$ and $\vecown{r}_{{\rm
    cac},q}^{(\rm{obj})}$ describe the coordinates of a given point in
the global and individual systems, respectively, $L_0$ and $L_q$ are
the length scales of the coordinate systems, $\vecown{t}_q$ is the
translation vector to the origin of the individual systems measured in
the global system, and the transformation matrices $\matown{Q}_q$ are
given by:
\beq
\label{eq:transmatq}
\matown{Q}_q=
\begin{pmatrix}
   \vecown{e}_x^{(q)} \cdot \vecown{e}_x^{(0)} & \vecown{e}_y^{(q)} \cdot \vecown{e}_x^{(0)} & \vecown{e}_z^{(q)} \cdot \vecown{e}_x^{(0)} \\
   \vecown{e}_x^{(q)} \cdot \vecown{e}_y^{(0)} & \vecown{e}_y^{(q)} \cdot \vecown{e}_y^{(0)} & \vecown{e}_z^{(q)} \cdot \vecown{e}_y^{(0)} \\
   \vecown{e}_x^{(q)} \cdot \vecown{e}_z^{(0)} & \vecown{e}_y^{(q)} \cdot \vecown{e}_z^{(0)} & \vecown{e}_z^{(q)} \cdot \vecown{e}_z^{(0)}
\end{pmatrix}
\,.
\eeq
The unit vectors $\vecown{e}_{x,y,z}^{(q)}$ of the individual
coordinate systems $\Sigma_{\rm{cac},q}^{(\rm{obj})}$ are described in
the global system $\Sigma_{\rm{cac},0}^{(\rm{obj})}$, and allow for
tilts of the individual objects within the global coordinate
system. In regions where individual coordinate systems are
overlapping, the local state of the gas needs to be described
consistently within all grids.

For a given observer's direction $\vecown{n}_0$ measured in the global
frame then, we compute the corresponding representations in the
individual coordinate systems, $\vecown{n}_q=\matown{Q}_q^{-1}\cdot
\vecown{n}_0$, and assign cylindrical coordinate systems to each of
the objects following the procedure outlined in
Sect.~\ref{sec:single_theory} by setting
$\vecown{e}_{\tilde{z}}^{(q)}=\vecown{n}_q$. Again, to keep track of
the different coordinate systems, we define the transformation from
the local ray-coordinates to the local object coordinates:
\beq
\label{eq:transform03}
\vecown{r}_{{\rm cac},q}^{(\rm{obj})} = \matown{A}_q \cdot \vecown{r}_{{\rm cac},q}^{(\rm{ray})}
\qquad
\Longleftrightarrow
\qquad
\vecown{r}_{{\rm cac},q}^{(\rm{ray})} = \matown{A}_q^{-1} \cdot \vecown{r}_{{\rm cac},q}^{(\rm{obj})} \,,
\eeq
with transformation matrix
\beq
\label{eq:transmata}
\matown{A}_q=
\begin{pmatrix}
\vecown{e}_{\tilde{x}}^{(\rm{q})} \cdot \vecown{e}_x & \vecown{e}_{\tilde{y}}^{(\rm{q})} \cdot \vecown{e}_x & \vecown{e}_{\tilde{z}}^{(\rm{q})} \cdot \vecown{e}_x \\
\vecown{e}_{\tilde{x}}^{(\rm{q})} \cdot \vecown{e}_y & \vecown{e}_{\tilde{y}}^{(\rm{q})} \cdot \vecown{e}_y & \vecown{e}_{\tilde{z}}^{(\rm{q})} \cdot \vecown{e}_y \\
\vecown{e}_{\tilde{x}}^{(\rm{q})} \cdot \vecown{e}_z & \vecown{e}_{\tilde{y}}^{(\rm{q})} \cdot \vecown{e}_z & \vecown{e}_{\tilde{z}}^{(\rm{q})} \cdot \vecown{e}_z
\end{pmatrix}
\,,
\eeq
and $\vecown{e}_x=(1,0,0)$, $\vecown{e}_y=(0,1,0)$,
$\vecown{e}_z=(0,0,1)$ the representation of the object's local
coordinate system within the corresponding basis. Accordingly, we also
define a global `cylindrical' coordinate system with
$\vecown{e}_{\tilde{z}}=\vecown{n}_0$, since the individual
cylindrical coordinate systems may overlap (\eg~during transits). The
flux integration, Eq.~\eqref{eq:fluxint}, is then performed in the
global cylindrical system, and needs to be adapted to avoid double
counting of overlapping areas. To this end, we consider the global
2D Delaunay-triangulation of all
$(p_{\indx{i}},\zeta_{\indx{j}})_q$ points from all individual
cylindrical coordinate systems by using the \textsc{GEOMPACK2}
library\footnote{\url{https://people.math.sc.edu/Burkardt/f_src/geompack2/geompack2.html}}
(\citealt{Joe1991}). The flux integration, Eq.~\eqref{eq:fluxint}, can
then be performed by numerically integrating the intensities over all
triangles.

With Eqs.~\eqref{eq:transform01}, \eqref{eq:transform02}, and
\eqref{eq:transform03} representing the formalism to transform
coordinates between the individual and global ray and object systems,
the basic algorithm is then defined as follows:
\begin{enumerate}
\item Definition of the individual coordinate systems for all objects
  $q$, that is length-scales $L_q$, translation vectors $\vecown{t}_q$,
  basis vectors (described in the global coordinate system)
  $\vecown{e}_{x,y,z}^{(\rm{q})}$, global velocities
  $\vecown{v}^{(\rm{q})}$ (corresponding to the orbital velocities of
  the objects), and calculation of the transformation matrices
  $\matown{Q}_q$ via Eq.~\eqref{eq:transmatq}.
\item Definition of the $(p,\zeta)_q$ grid for all objects $q$, and
  calculation of the transformation matrices between ray and object
  coordinate systems $\matown{A}_q$ via Eq.~\eqref{eq:transmata}.
\item Transformation of all $(p,\zeta)_q$ coordinates to the global
  ray-coordinate system using Eqs.~\eqref{eq:transform01} and
  \eqref{eq:transform03} to obtain the corresponding
  $(\tilde{x},\tilde{y})$ coordinates in the global ray coordinate
  system.
\item Calculation of the 2D triangulation in the global ray system.
\item Back-transformation of each $(\tilde{x},\tilde{y})$ coordinate
  to all individual coordinate systems via Eqs.~\eqref{eq:transform02}
  and \eqref{eq:transform03}.
\item Step 2 of the single-star algorithm: Setting the
  $\tilde{z}_q$ grid along a ray within each individual coordinate
  system following Eq.~\eqref{eq:zgrid}.
\item Step 3 of the single-star algorithm: Transformation of all
  $(\tilde{x}_{\indx{i}},\tilde{y}_{\indx{j}},\tilde{z}_{\indx{k}})_q$
  coordinates to the individual spherical systems and interpolation of
  all required physical quantities (\ie~opacities, source functions,
  velocity fields) onto the ray.
\item Transformation of the individual rays to the global ray system
  via Eqs.~\eqref{eq:transform01} and \eqref{eq:transform03}.
\item Merging all individual rays to one single ray for a given
  $(\tilde{x},\tilde{y})$ point (see Fig.~\ref{fig:transform01}).
\item Step 4 of the single-star algorithm: Solving the
  discretized form of the radiative transfer equation,
  Eq.~\eqref{eq:formal_solution}, along the ray.
\item Step 5 of the single-star algorithm: Integration of
  emerging specific intensities over the triangulated area in
  Eq.~\eqref{eq:fluxint} to obtain the flux at a specific frequency
  bin.
\item Step 6 of the single-star algorithm: Restarting the
  procedure at step 6 for the next frequency bin.
\end{enumerate}

With this algorithm, we are able to calculate synthetic line profiles
for any given orbital configuration and any given state of the
surrounding material.  We emphasize that this procedure is independent
of the length scales of the individual systems, and automatically
accounts for transits. Thus, one could also analyse, for example,
planetary atmospheres or complex jet configurations as frequently
found in post-asymptotic giant branch binary systems
(\eg~\citealt{Bollen2020}). Moreover, this algorithm can be extended
to multiple systems, with the computation time typically scaling
linearly with the number of objects, and a quadratic scaling found
only for complex situations (\eg~when all involved objects are
transiting).
%
%
\section{LB-1}
\label{sec:lb1}
With the above described algorithm, we can readily tackle binary
systems in 3D (see also Appendix~\ref{app:tests} for some test
calculations of the code when applied to single-star models). In this
section we model the \Ha~line of the binary (or multiple) system
\mbox{LB-1} as a first application, focussing on the strB+Be scenario
by \cite{Shenar2020} and on the original B+BH scenario (though with a
revised mass for the B star by \citealt{SimonDiaz2020}). For both
hypotheses, we adopted the stellar parameters as obtained by these
authors. For the B star in the B+BH scenario, \cite{SimonDiaz2020}
applied the 1D spherically symmetric NLTE code \textsc{FASTWIND} to
deduce the stellar parameters. Similarly, \cite{Shenar2020} applied
the 1D spherically-symmetric NLTE code \textsc{PoWR}
(\citealt{Hamann2003}, \citealt{Sander2017b}) to the disentangled
spectrum of the Be-star in the strB+Be scenario, while relying on the
\textsc{Grid Search in Stellar Parameters} tool (\textsc{GSSP},
\citealt{Tkachenko2015}) with synthetic spectra derived from 1D
plane-parallel LTE modelling using the \textsc{synthv} radiative
transfer code (\citealt{Tsymbal1996}) and a grid of \textsc{LLmodel}
atmospheres (\citealt{Shulyak2004}) when analysing the stripped
star. The updated stellar parameters by \cite{Lennon2021} using the 1D
plane-parallel NLTE code \textsc{TLUSTY} (\citealt{Hubeny88},
\citealt{Hubeny95}) are typically in a similar range (see
Table~\ref{tab:params_lb1}). These studies relied on investigating
photospheric lines that are (at most) very weakly contaminated by the
disc (in contrast to \Ha, for instance), so that the adopted stellar
parameters should be good representatives of the underlying stars.
With the given stellar parameters then, we ‘only’ need to define a
suitable disc model for the BH disc and the Be-star disc to calculate
synthetic \Ha~line profiles.
  
To this end, we have set up an axisymmetric analytical model for the disc
following \cite{Kee2018a} (see also \citealt{Carciofi2006}) which is
based on a prescribed density stratification in the equatorial plane
by a parameterised power-law, and assumes hydrostatic equilibrium in
the vertical direction. Then, the density in the disc is:
\beq
\label{eq:rhodisc}
\rho \left( r,\clatitude \right) = \rho_0 \left( \dfrac{r
  \sin\clatitude}{\Rstar}\right)^{-\beta_{\rm D}} \exp \left[-\dfrac{G
    \Mstar}{\vsound^2}\left(\dfrac{1}{r \sin \clatitude}-\dfrac{1}{r}
  \right) \right] \,,
\eeq
with $r,\clatitude$ the radial and co-latitudinal coordinates,
$\Mstar$ the stellar or BH mass, $\Rstar$ the stellar radius or the
minimum radius of the BH disc, and $\rho_0$ and $\beta_{\rm D}$ the
base density and power-law index for the density stratification. When
$\beta_{\rm D}=15/8$, this formulation is equivalent to the standard
$\alpha$-accretion disc model by \cite{ShakuraSunyaev1973} at large
distances from the BH, with the $\alpha$-prescription and accretion
rate hidden\footnote{Following the formulation by
      \cite{frankbook2002}, the base density can be derived from their
      Eq.~5.49 at large distances from the accreting object, yielding
      $\rho_0 = 3.1 \cdot 10^{-8} \alpha^{-7/10} \dot{M}_{16}^{11/20}
      m_1^{5/8}R_{\ast,10}^{-15/8}$, with $\dot{M}_{16}$ the
      mass-accretion rate in $10^{16}\,{\rm g/s}$, $m_1$ the mass of
      the central object in $\msun$, and $R_{\ast,10}$ the inner disc
      radius in units of $10^{10}\,{\rm cm}$.} within the parameter
$\rho_o$. As such, $\rho_0$ is related to the mass flux through the disc, and
  $\beta_D$ can be considered as a combined proxy for the disc
  viscosity, scale height, and temperature as a function of radius
  (within both a BH accretion or a Be-star decretion disc). For
Be-stars, typical values are of the order of $\beta_{\rm D}\in[1.5,4]$
and $\rho_0\in 5 \cdot[10^{-10},10^{-13}]\,\gcmc$ (based on
  observations and \Ha~line fitting, \eg~\citealt{Silaj2010}, see
also the review by \citealt{Rivinius2013}). With the sound speed,
$\vsound=\sqrt{\kboltz T / \mu m_p}$, evaluated here for a typical
temperature $T=10\,\kK$ and a mean molecular weight $\mu=0.6$, the
disc density is specified for given stellar parameters and the input
parameters $\rho_0$, $\beta_{\rm D}$.

Further, we assumed Keplerian rotation of the disc in the orbital plane
of the binary system, with the azimuthal component of the velocity
field given as
\beq
\label{eq:vrot}
v_\azimuth \left( r,\clatitude \right) = \sqrt{\dfrac{G\Mstar}{r\sin \clatitude}} \,.
\eeq
Since the temperature stratification crucially depends on the type of
the disc (BH-accretion disc or Be-star disc), the formulation
corresponding to each individual model will be described in
Sects.~\ref{subsec:lb1_bh} and \ref{subsec:lb1a}, respectively. To
calculate \Ha~opacities and source functions, we applied
Eqs.~\eqref{eq:defopal} and \eqref{eq:defsline}, with occupation
numbers calculated from Saha-Boltzmann statistics assuming full
ionization.

For such disc models, an overarching goal is to qualitatively
reproduce the observed \Ha~line profiles (see
Fig.~\ref{fig:profiles_best}), together with the corresponding
dynamical spectrum and the radial velocity curves with a
semi-amplitude for the primary and secondary of $K_1\approx 53\,\kms$
and $K_2\approx[6,13]\,\kms$, respectively. For the secondary object,
the interpretation of the radial velocity curve crucially depends on
the applied method, with essentially two choices at hand. Firstly, the
semi-amplitude can be calculated from the barycentric method, where
the \Ha~line centre at each phase is determined by considering the
line wings up to a certain flux level. Using this method,
\cite{Liu2019} estimated $K_2^{\rm (H_\alpha)} \approx
6\,\kms$. However, as shown by \cite{AbdulMasih2020}, an absorption
component of the primary object could increase the apparent motion of
the line wings, thus overestimating the true orbital
velocities. Following this argumentation, an emission component of the
primary object would underestimate the true orbital velocities when
measured with the barycentric method.  Alternatively, the orbital
solution could be estimated by applying spectral disentangling
techniques, with corresponding results expected to represent the
actual orbital motion of the objects. Using this method,
\cite{Shenar2020} measured $K_2^{\rm (true)}\approx 11\,\kms$ for
\mbox{LB-1}, in good agreement with a detailed investigation of
near-infrared emission lines by \cite{Liu2020} estimating $K_2^{\rm
  (true)}\in [8,13]\,\kms$. The differences between the $K_2$ values
as obtained from the barycentric method and the disentangled spectra
might then be explained by an emission component attached to the
primary object.  Our modelling then should ideally reproduce the
$K_2^{\rm (H_\alpha)}\approx6\,\kms$ semi-amplitude when applying the
barycentric method to the synthetic \Ha~lines, while also yielding
$K_2^{\rm(true)}\approx 11\,\kms$ from the actual orbit of the system.

Since the eccentricity is small ($e=0.03\pm 0.01$, see
\citealt{Liu2019}), we assumed a circular orbit. For given masses of
the individual objects, $M_1$, $M_2$, and an observed orbital period
$P\approx 79\,{\rm d}$ then, the orbit is obtained from the two-body
problem:
\beqa
\label{eq:orbit_binary0}
a &=& \left[\dfrac{G \left(M_1 + M_2 \right) P^2}{4 \pi^2}
  \right]^{1/3} \\
\label{eq:orbit_binary1}
a_1 &=& \dfrac{M_2}{M_1+M_2} a \qquad v_1 = \dfrac{M_2}{M_1 + M_2}
\dfrac{2 \pi a}{P} \\
\label{eq:orbit_binary2}
a_2 &=& \dfrac{M_1}{M_1+M_2} a \qquad v_2 = \dfrac{M_1}{M_1 + M_2}
\dfrac{2 \pi a}{P}\,,
\eeqa
with $a$ the binary separation, $a_1$, $a_2$ the distance of the
objects to the centre of mass, and $v_1$ and $v_2$ the absolute
orbital velocities. Thus, for the observed radial velocity curve(s)
and period, and assuming the individual masses of the objects to be
given, the inclination $i$ is fixed within our calculations in order
to reproduce the semi-amplitude of the radial velocity curve for the
primary object by $v_1 \cdot \sin(i) = K_1$.  For our best models, the
orbital parameters are summarized in Table~\ref{tab:params_lb1_orbit}.
\subsection{B-star and BH-disc (B+BH) scenario}
\label{subsec:lb1_bh}
\begin{figure*}[ht]
\resizebox{\hsize}{!}{
   \resizebox{0.25\hsize}{!}{\includegraphics{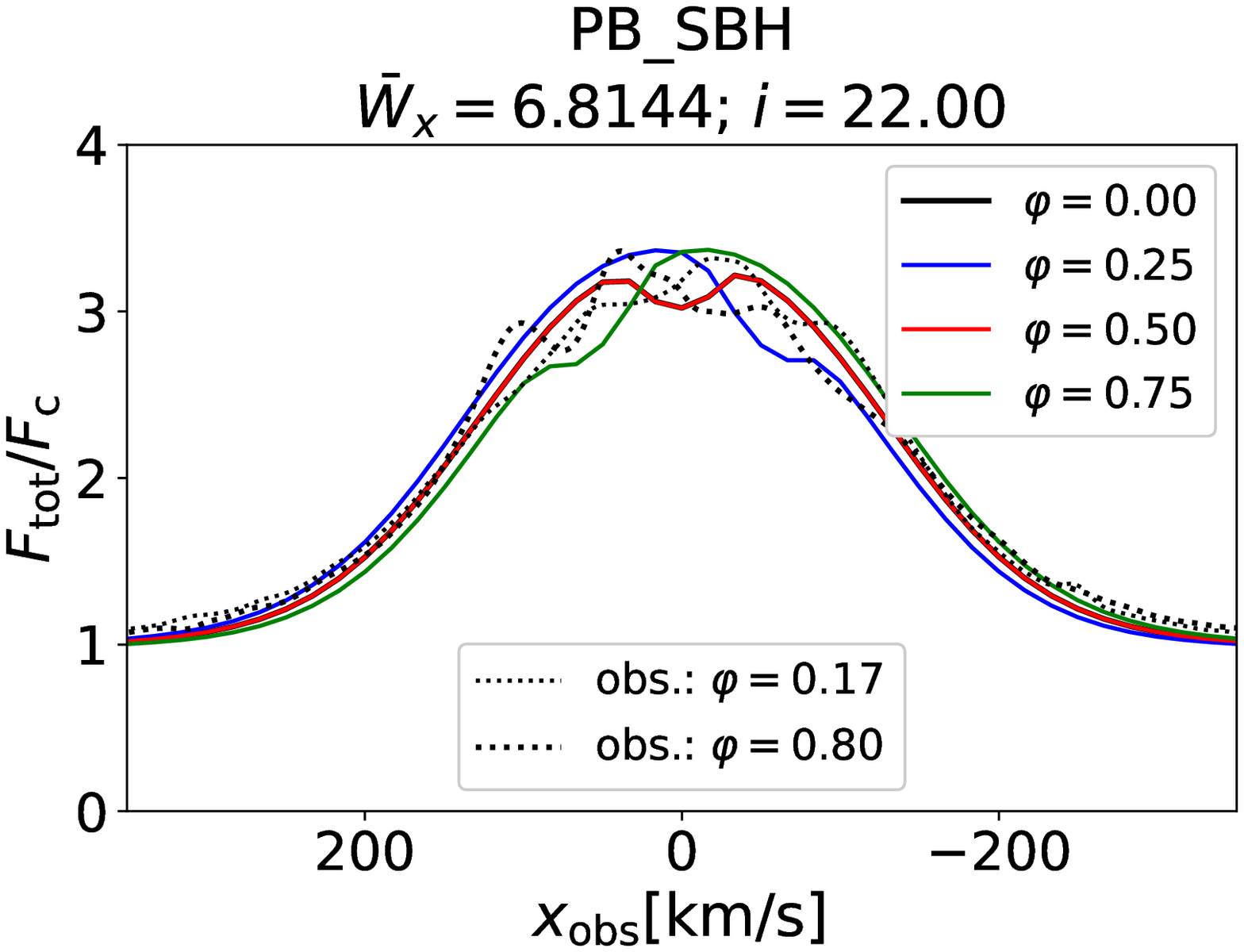}}
   \resizebox{0.25\hsize}{!}{\includegraphics{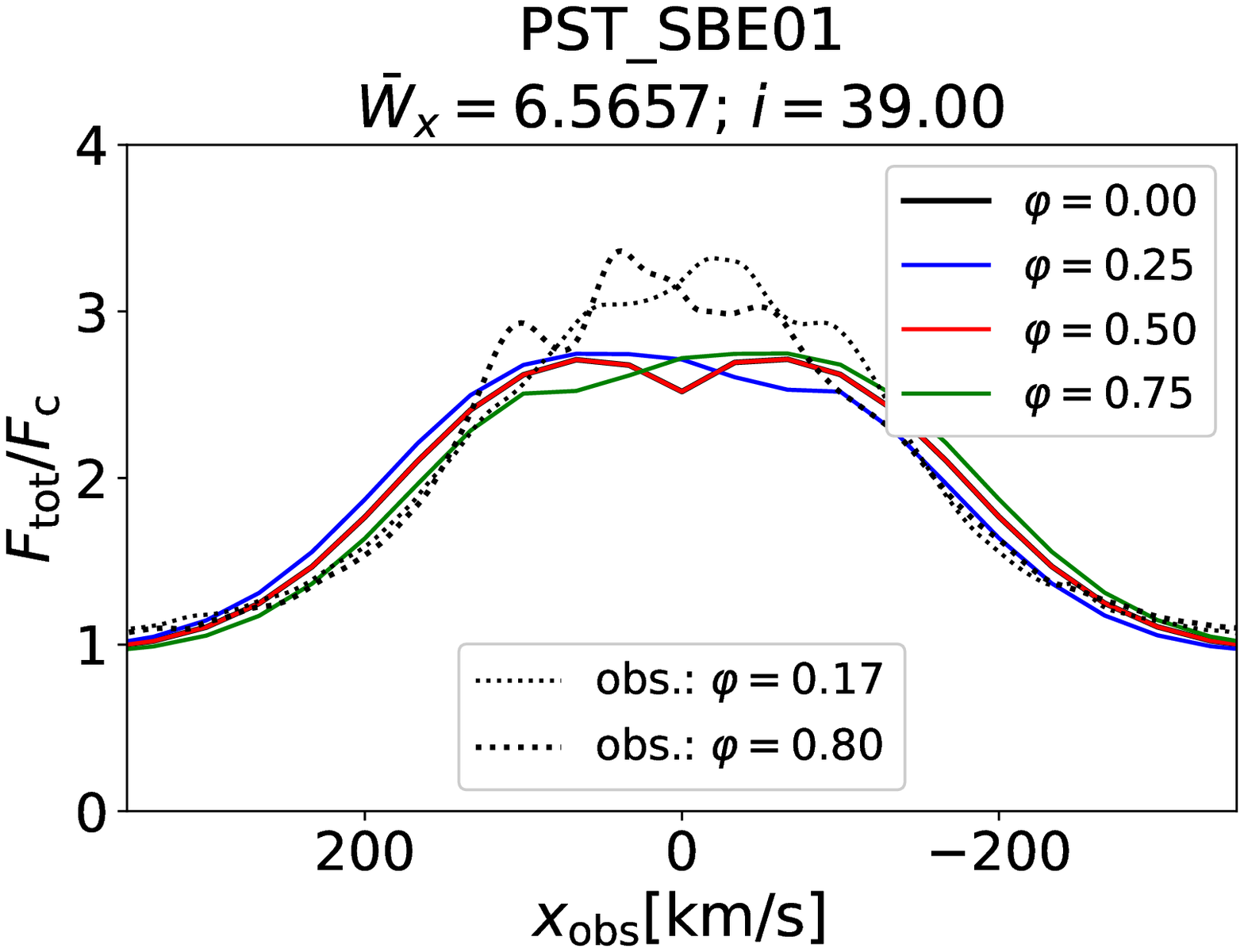}}
   \resizebox{0.25\hsize}{!}{\includegraphics{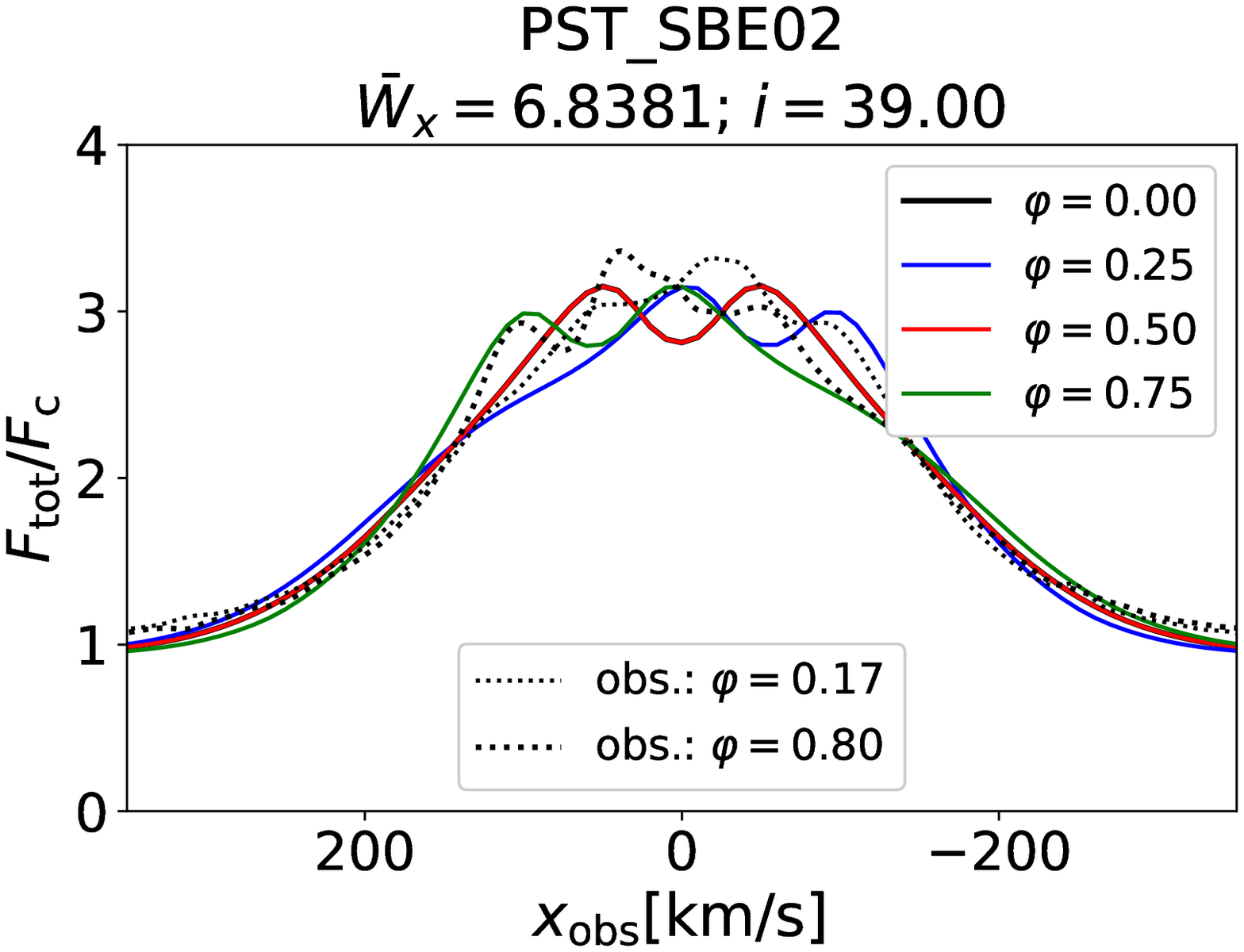}}
   \resizebox{0.25\hsize}{!}{\includegraphics{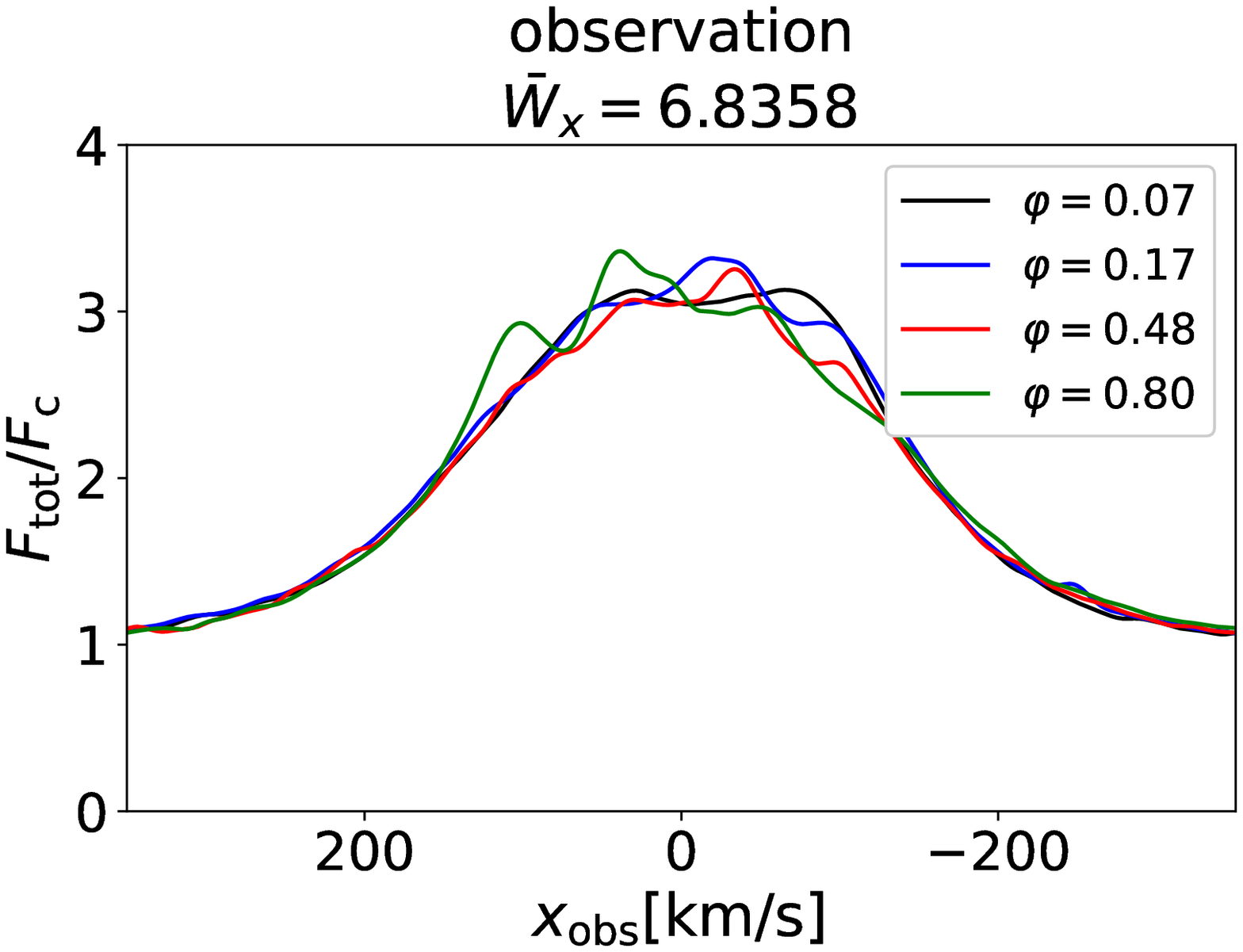}}    
}
\resizebox{\hsize}{!}{
   \resizebox{0.25\hsize}{!}{\includegraphics{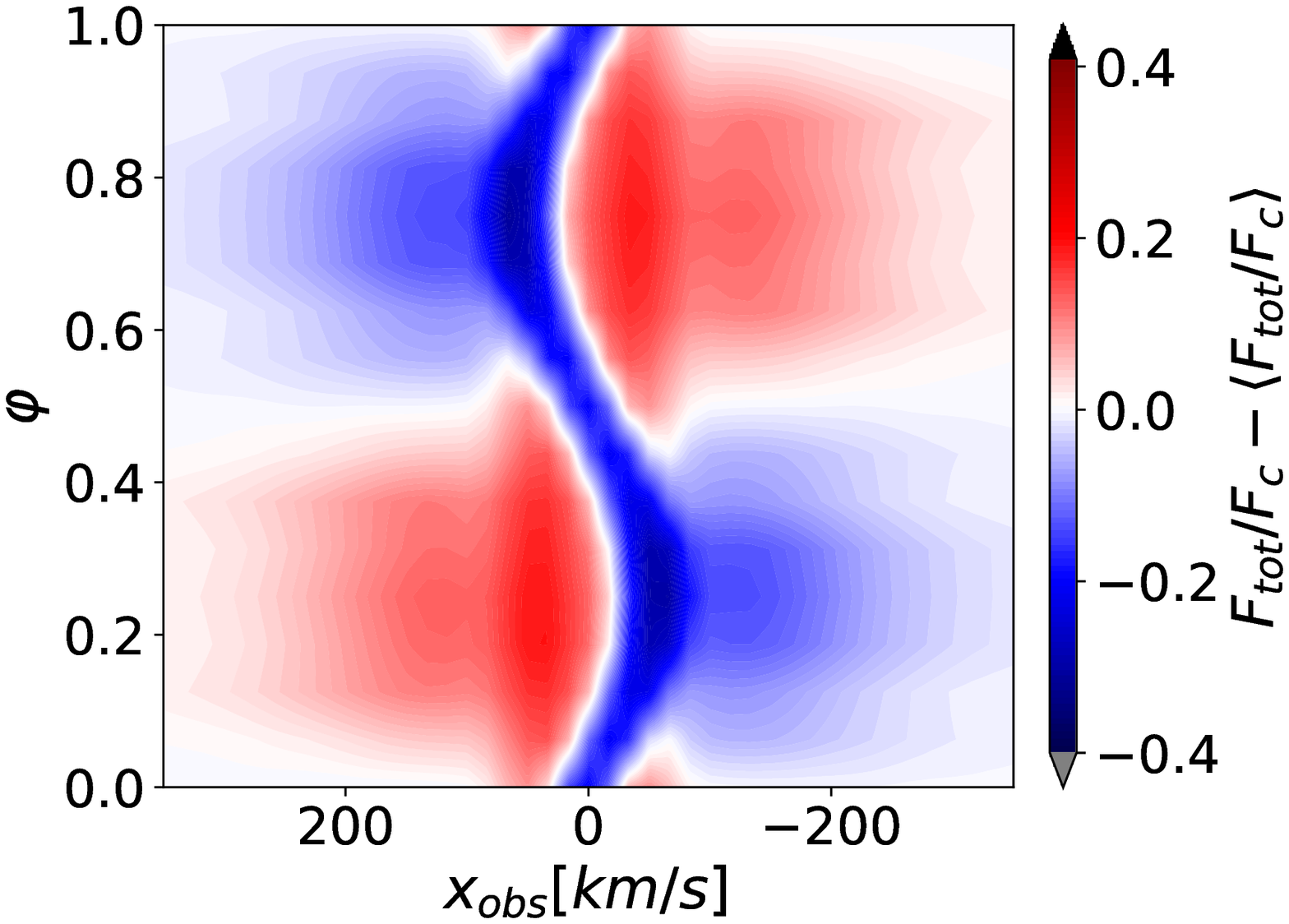}}
   \resizebox{0.25\hsize}{!}{\includegraphics{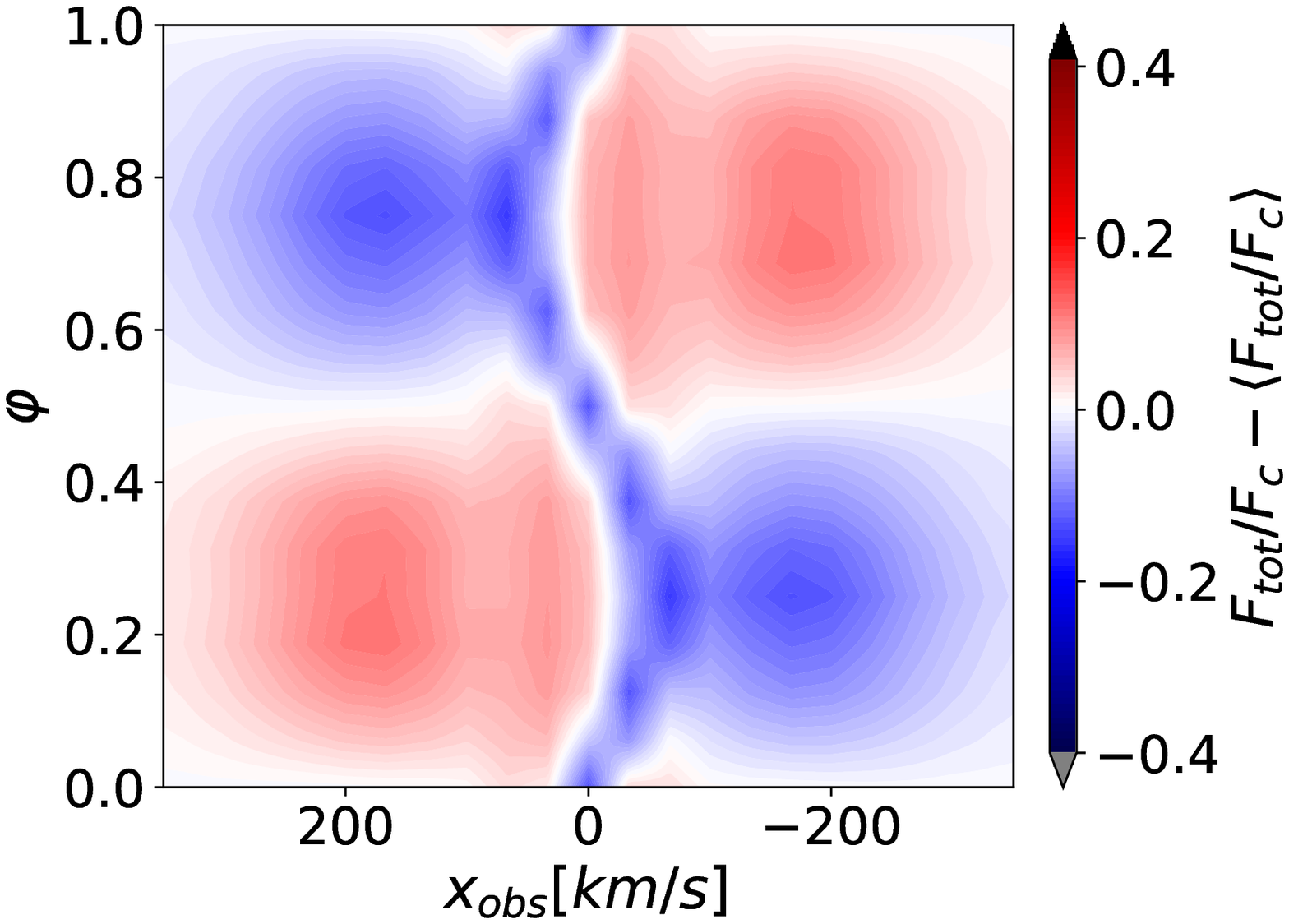}}
   \resizebox{0.25\hsize}{!}{\includegraphics{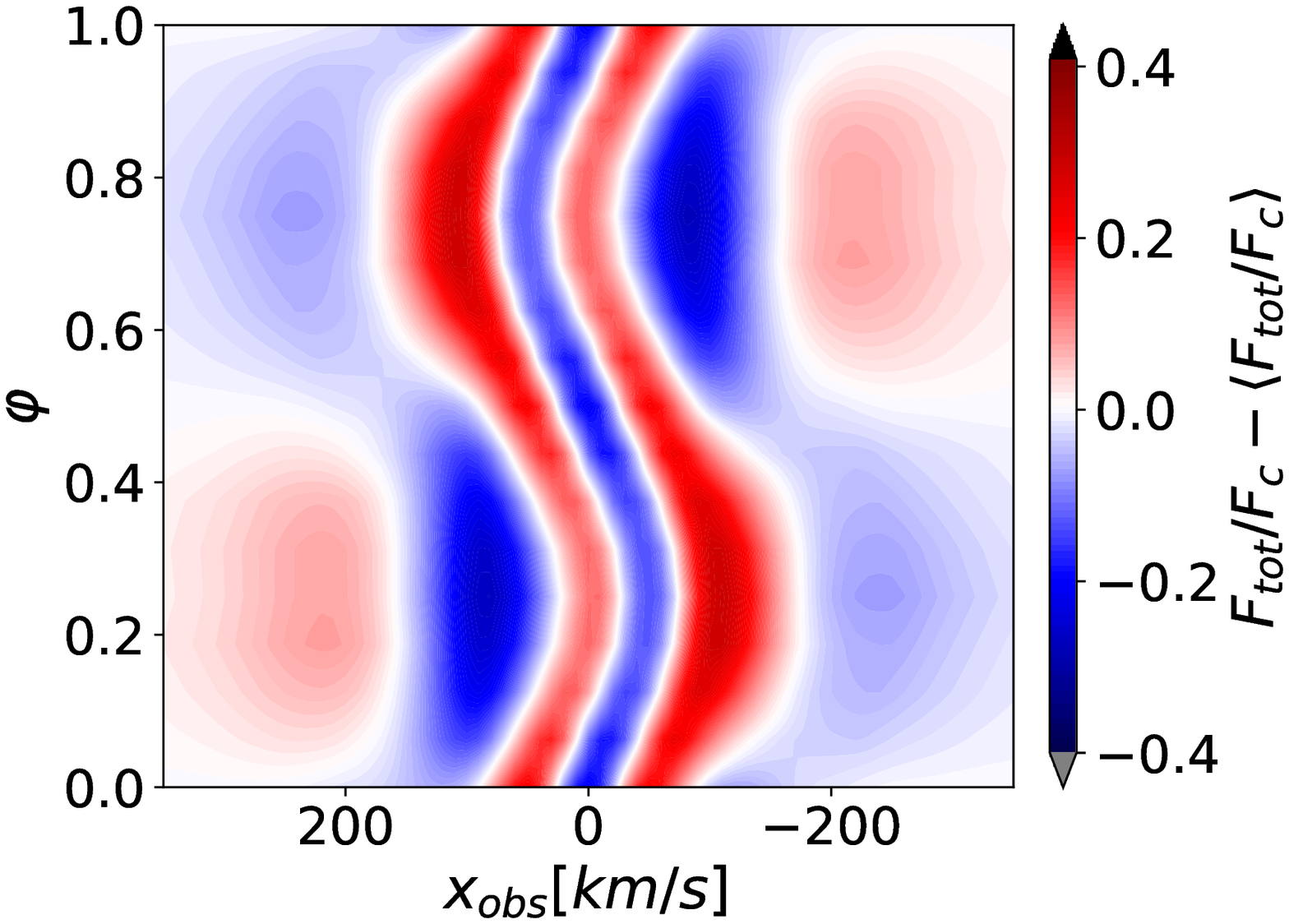}}
   \resizebox{0.25\hsize}{!}{\includegraphics{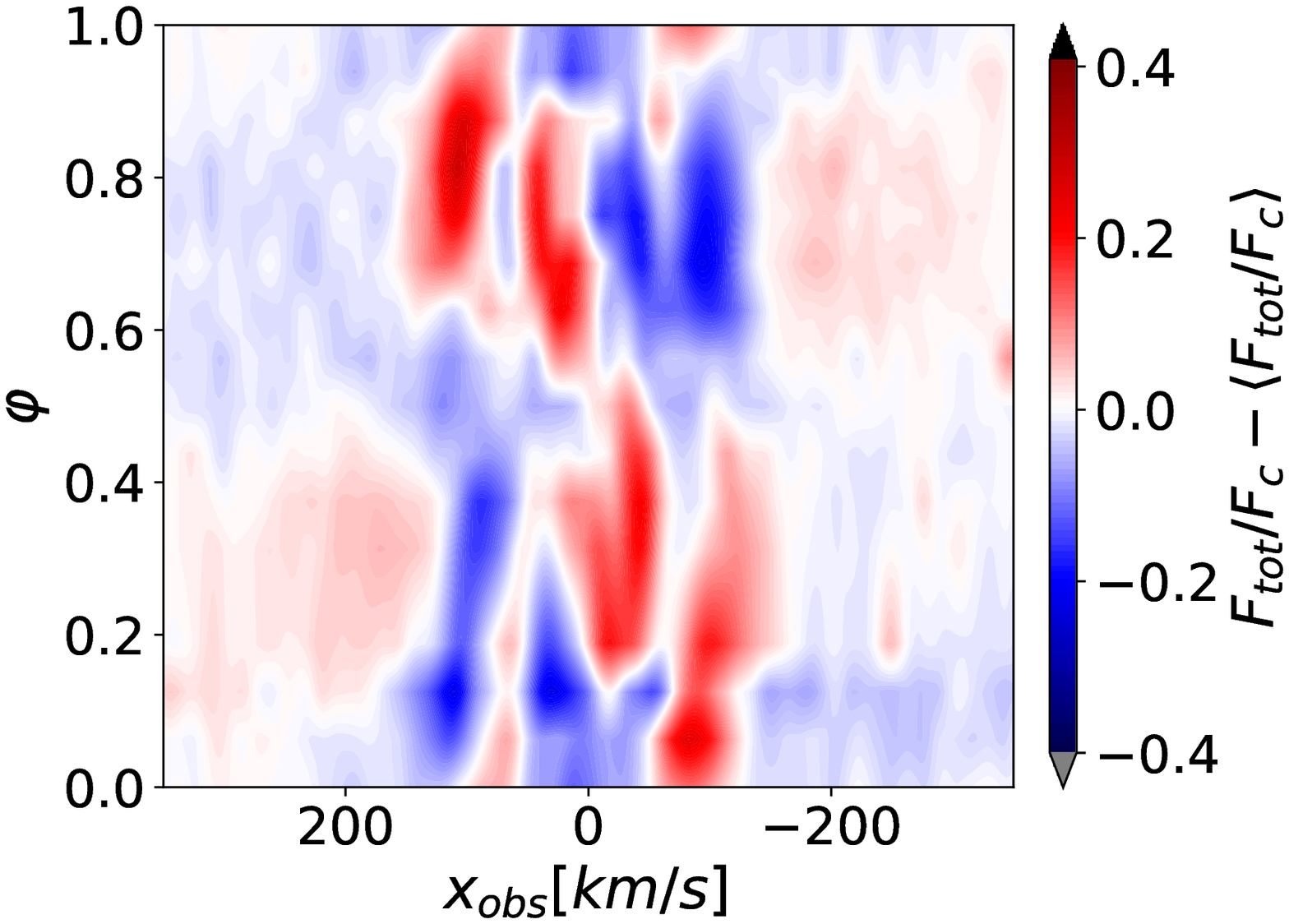}}    
}
\resizebox{\hsize}{!}{
   \resizebox{0.25\hsize}{!}{\includegraphics{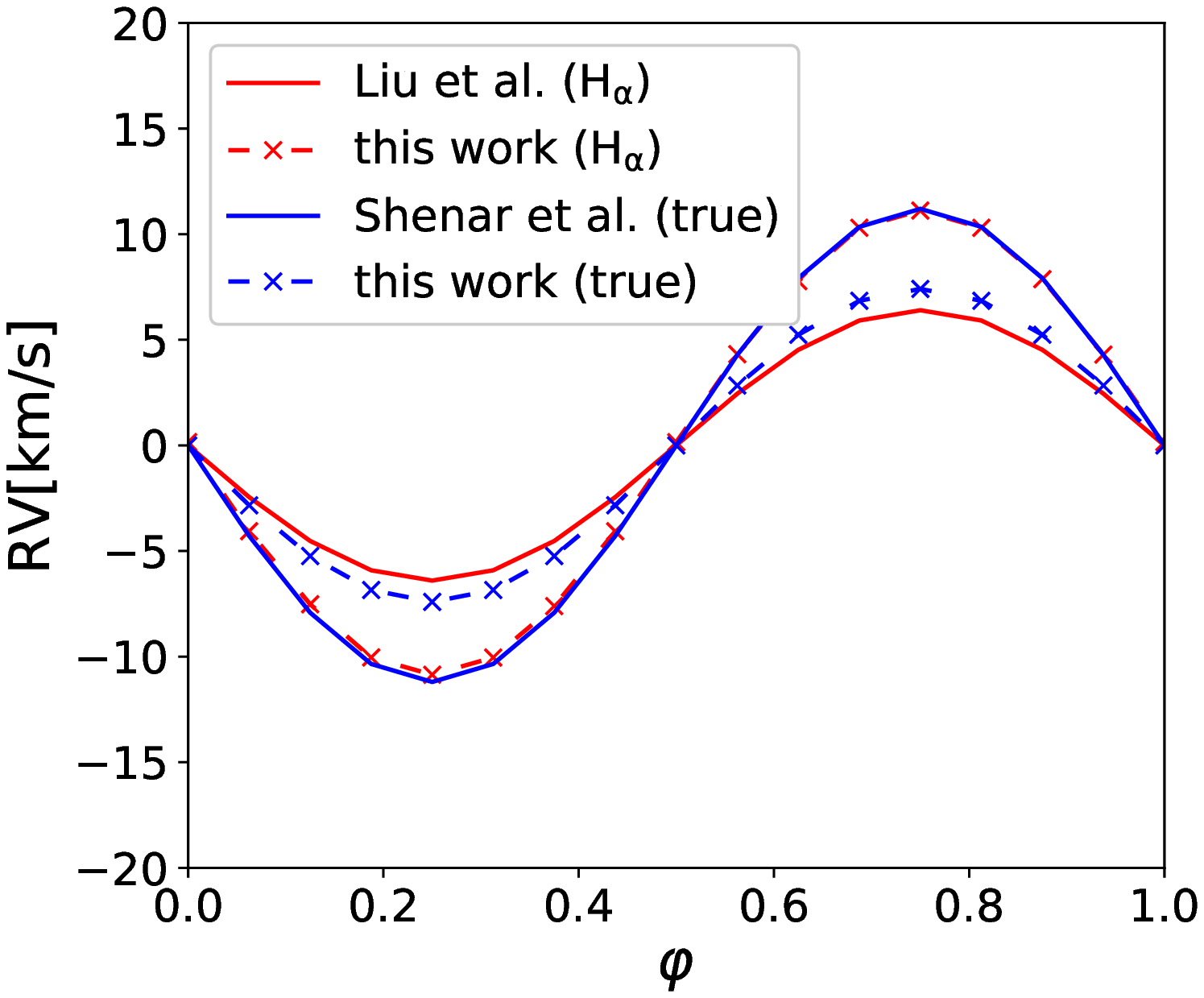}}
   \resizebox{0.25\hsize}{!}{\includegraphics{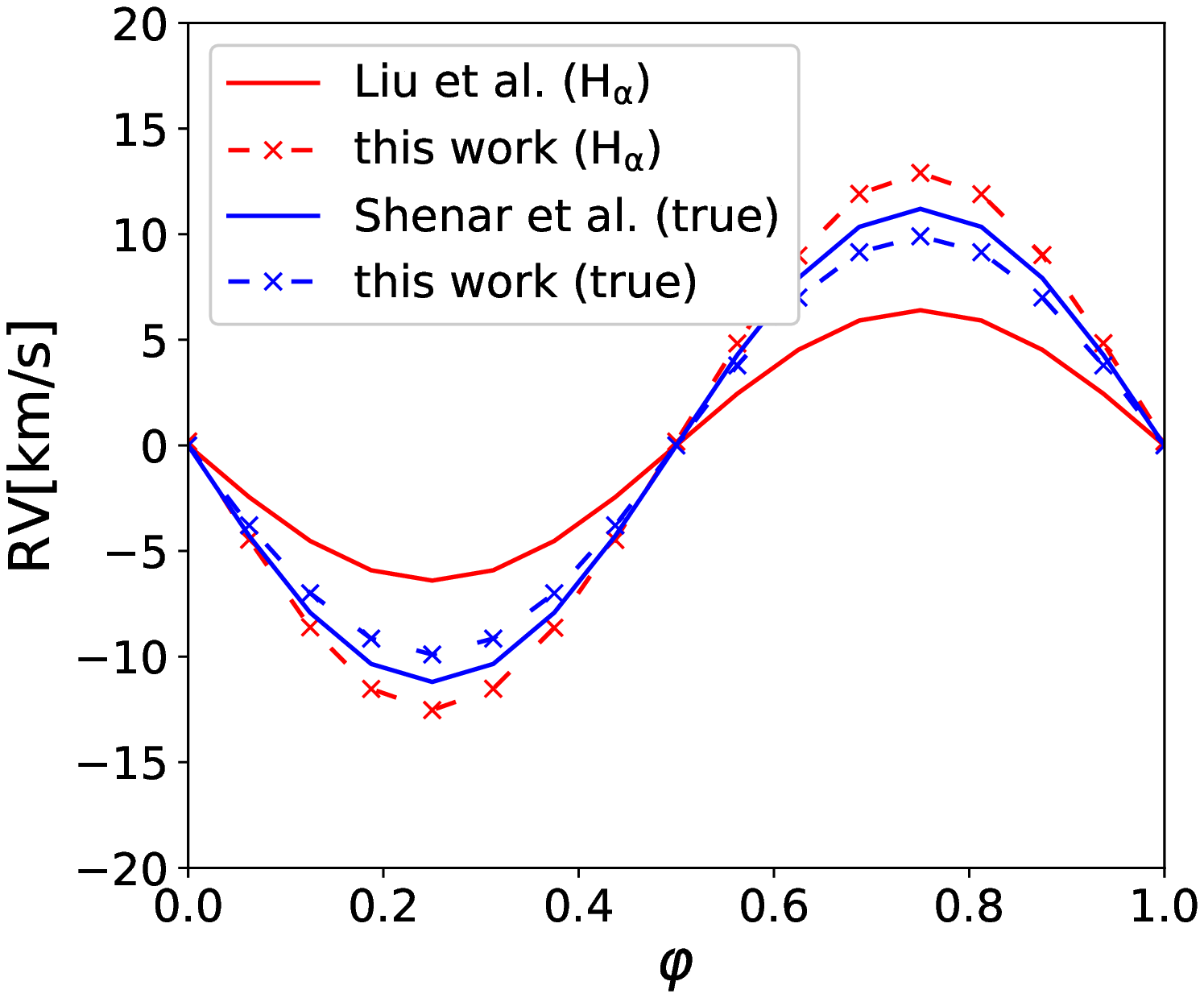}}
   \resizebox{0.25\hsize}{!}{\includegraphics{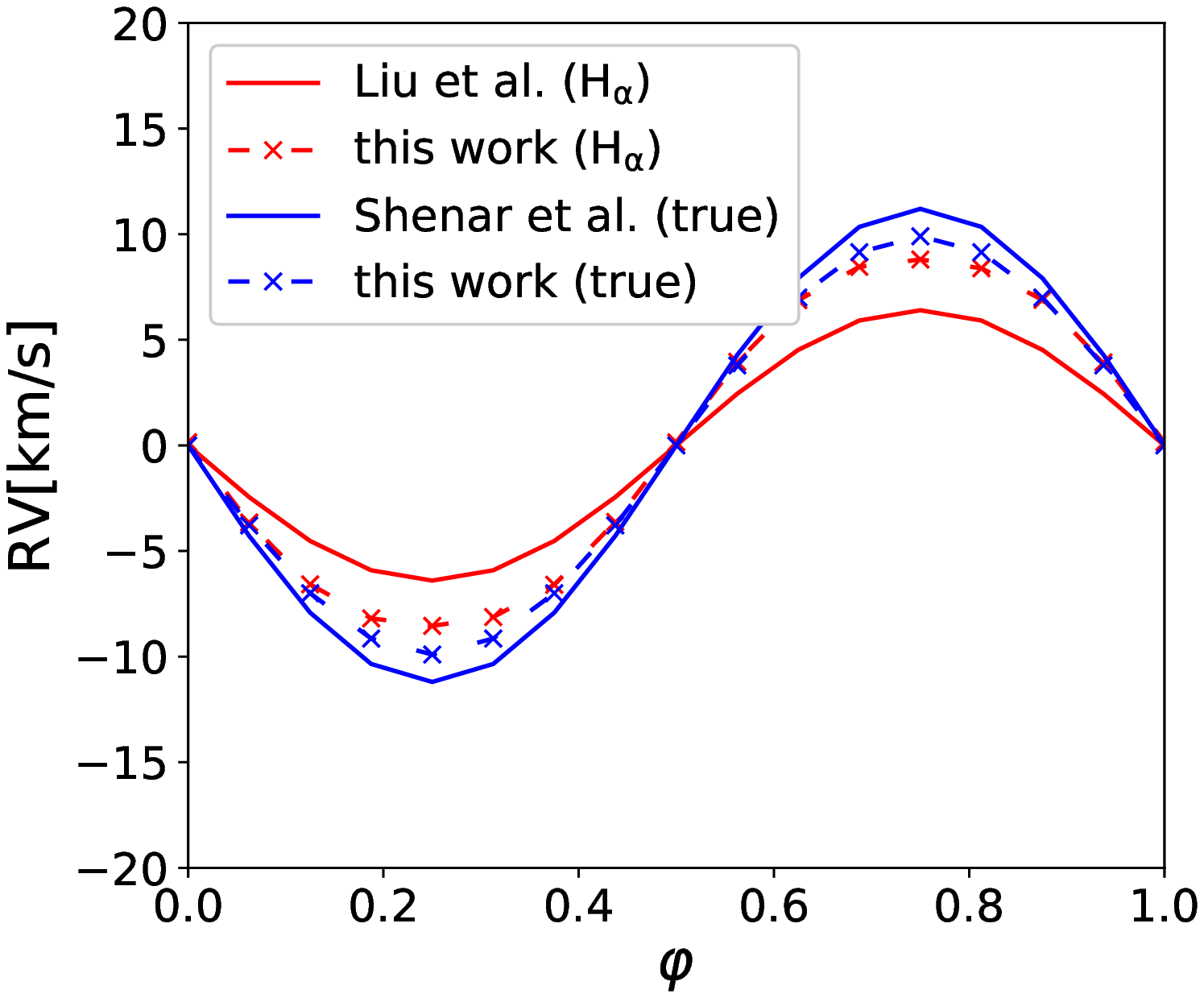}}
   \resizebox{0.25\hsize}{!}{\includegraphics{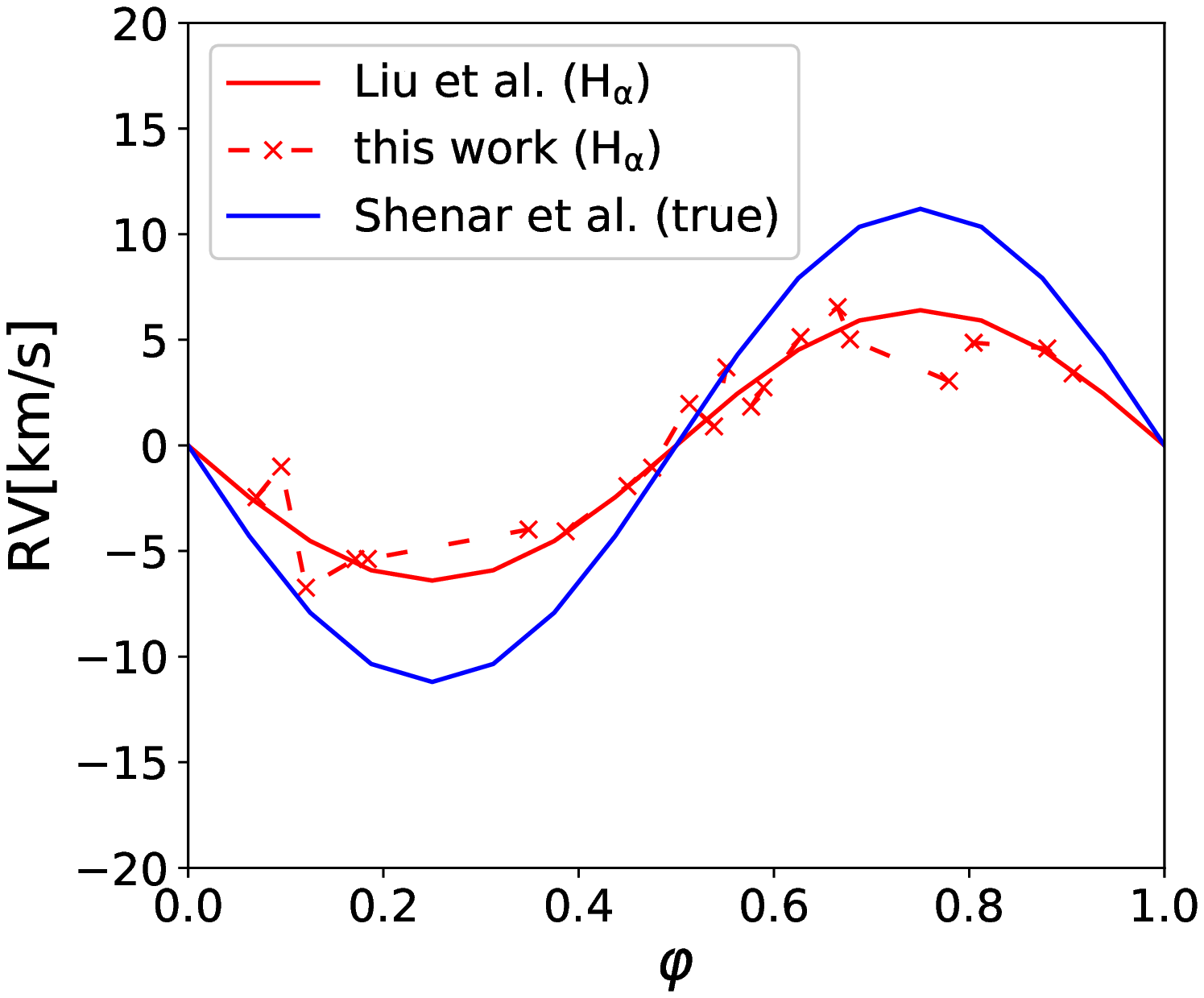}}  
}
  \caption[Small caption.]{\Ha~line profiles at different phases
    $\varphi$ (top row) with corresponding mean-subtracted dynamical
    spectra (middle row) and radial velocity curves (bottom row) of
    the BH or strB companion. The $x_{\rm obs}$-axis describes the frequency
    shift from line centre expressed in velocity space.  The first
    three columns to the left display the solution for the three
    different models, PB\_SBH (primary: B star, secondary: BH),
    PST\_SBE01 (primary: stripped star, secondary: Be-star, model 1),
    PST\_SBE02 (primary: stripped star, secondary: Be-star, model 2),
    with \Ha~line profiles at two distinct phases from the actual
    observations indicated by the black dashed and dotted lines. The
    right column additionally displays the corresponding observed (see
    \citealt{Shenar2020}) \Ha~line profiles, mean-subtracted dynamical
    spectrum and radial velocity curves of \mbox{LB-1}, where we distinguish
    between the true ones (\eg~from using disentangling methods)
    indicated in blue and those obtained from the barycentric method
    applied to the \Ha~line wings (red). To reduce the noise, all
    observed line profiles have been convolved with a Gaussian filter
    of width $10 \,\kms$.  The bottom panels display the radial
    velocity curves of the secondary when calculated from the
    barycentric method applied to the \Ha~line~profile wings (red
    crosses) and when using the true orbital parameters (blue
    crosses). Additionally, we display the corresponding radial
    velocity curves as found by \cite{Shenar2020} and \cite{Liu2019}
    using disentangling techniques (blue solid line) and the
    barycentric method (red solid lines), respectively.  At the top,
    the (phase-averaged) equivalent width evaluated in velocity space
    in units of $100\,\kms$ and the inclination used within the
  synthetic-spectra calculations are indicated.}
\label{fig:profiles_best}
\end{figure*}
\begin{figure}[ht]
\resizebox{\hsize}{!}{\includegraphics{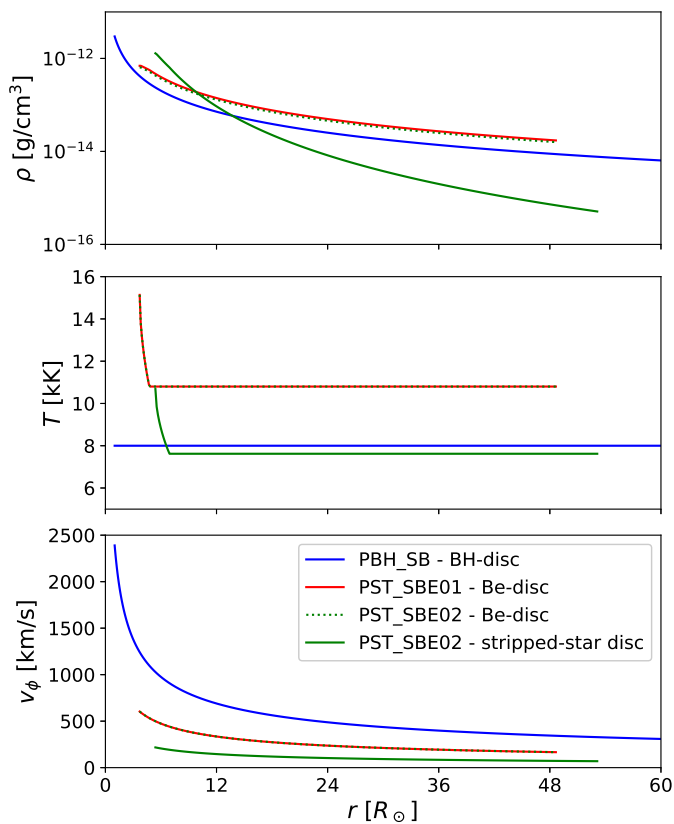}}
\caption[Small caption.]{Disc models in the equatorial plane
    corresponding to our best-fit parameters of the LB-1 system (see
    Table~\ref{tab:params_lb1_fit}). The top and middle panels display
    the radial density and temperature stratification, respectively,
    and the azimuthal velocity component is shown in the bottom
    panel. The blue and red lines indicate the BH accretion disc
    (model PBH\_SB) and the Be-star disc within the single-disc model
    (model PST\_SBE01), respectively. For model PST\_SBE02 (containing
    two discs), the Be-star disc is indicated by the green dotted line
    (partially overlapping with the red solid line from model
    PST\_SBE01), and the stripped-star disc is indicated by the green
    solid line.}
\label{fig:models_best}
\end{figure}
%
%
%
%
%
\begin{table*}
\setlength{\tabcolsep}{1.28mm}
\begin{center}
\caption{Stellar parameters for the B+BH and strB+Be scenario
    used for our simulations of the \mbox{LB-1} system, and as found
    in the literature.}
\label{tab:params_lb1}
%
%
\begin{tabular}{ccccccccc}
\hline
\hline
\noalign{\vskip 0.55mm}
\multicolumn{1}{c}{} &
\multicolumn{1}{c}{} &
\multicolumn{3}{c}{B+BH} &
\multicolumn{1}{c}{} &
\multicolumn{3}{c}{strB+Be}
\\
\cline{3-5}
\cline{7-9}
\noalign{\vskip 0.55mm}
\multicolumn{1}{c}{} &
\multicolumn{1}{c}{reference} &
\multicolumn{1}{c}{this work} &
\multicolumn{1}{c}{(1)} &
\multicolumn{1}{c}{(2)} &
\multicolumn{1}{c}{} &
\multicolumn{1}{c}{this work} &
\multicolumn{1}{c}{(3)} &
\multicolumn{1}{c}{(2)}
\\
\cline{3-5}
\cline{7-9}
\noalign{\vskip 0.55mm}
\parbox[t]{2mm}{\multirow{7}{*}{\rotatebox[origin=c]{90}{primary}}} &
\multicolumn{1}{c}{type} &
\multicolumn{1}{c}{B star} &
\multicolumn{1}{c}{B star} &
\multicolumn{1}{c}{B star} &
\multicolumn{1}{c}{} &
\multicolumn{1}{c}{stripped star} &
\multicolumn{1}{c}{stripped star} &
\multicolumn{1}{c}{stripped star}
\\
\noalign{\vskip 0.55mm}
\multicolumn{1}{c}{} &  
\multicolumn{1}{c}{$T_{\rm eff}^{\rm (1)}\,[\kK]$} &
\multicolumn{1}{c}{14} &
\multicolumn{1}{c}{$\in \left[13.5,14.5\right]$} &
\multicolumn{1}{c}{$\in \left[15.0,15.6\right]$} &
\multicolumn{1}{c}{} &
\multicolumn{1}{c}{12.7} &
\multicolumn{1}{c}{$12.7$} &
\multicolumn{1}{c}{$\in \left[12.4,12.6\right]$}
\\
\noalign{\vskip 0.55mm}
\multicolumn{1}{c}{} &  
\multicolumn{1}{c}{$\Rstar^{(1)} \,[\rsun]$} &
\multicolumn{1}{c}{5.3} &
\multicolumn{1}{c}{$\in \left[4.1,6.5\right]$} &
\multicolumn{1}{c}{$\in \left[5.4,6.7\right]$} &
\multicolumn{1}{c}{} &
\multicolumn{1}{c}{5.4} &
\multicolumn{1}{c}{$5.4$} &
\multicolumn{1}{c}{$\in \left[4.4,5.3\right]$}
\\
\noalign{\vskip 0.55mm}
\multicolumn{1}{c}{} &  
\multicolumn{1}{c}{$\Mstar^{(1)} \,[\msun]$} &
\multicolumn{1}{c}{4.2} &
\multicolumn{1}{c}{$\in \left[1.3,5.5\right]$} &
\multicolumn{1}{c}{$\in \left[3.8,7.0\right]$} &
\multicolumn{1}{c}{} &
\multicolumn{1}{c}{1.3} &
\multicolumn{1}{c}{$\in \left[1.1,1.9\right]$} &
\multicolumn{1}{c}{$\in \left[0.5,1.3\right]$} 
\\
\noalign{\vskip 0.55mm}
\multicolumn{1}{c}{} &  
\multicolumn{1}{c}{$\logg^{(1)}$} &
\multicolumn{1}{c}{3.5} &
\multicolumn{1}{c}{$\in \left[3.35,3.65\right]$} &
\multicolumn{1}{c}{$\in \left[3.4,3.8\right]$} &
\multicolumn{1}{c}{} &
\multicolumn{1}{c}{3.1} &
\multicolumn{1}{c}{$3.0$} &
\multicolumn{1}{c}{$\in \left[2.8,3.2\right]$}
\\
\noalign{\vskip 0.55mm}
\multicolumn{1}{c}{} &  
\multicolumn{1}{c}{$\vrot^{(1)}\,[\kms]$} &
\multicolumn{1}{c}{21} &
\multicolumn{1}{c}{$<27^{\rm{(b)}}$} &
\multicolumn{1}{c}{$21^{\rm{(c)}}$} &
\multicolumn{1}{c}{} &
\multicolumn{1}{c}{11} &
\multicolumn{1}{c}{$\in [8,14]$} &
\multicolumn{1}{c}{$11$}
\\
\noalign{\vskip 0.55mm}
\multicolumn{1}{c}{} &  
\multicolumn{1}{c}{$\yhe^{(1)}$} &
\multicolumn{1}{c}{0.1} &
\multicolumn{1}{c}{$0.1$} &
\multicolumn{1}{c}{$0.1$} &
\multicolumn{1}{c}{} &
\multicolumn{1}{c}{0.2} &
\multicolumn{1}{c}{$0.21$} &
\multicolumn{1}{c}{$0.2$}
\\
\cline{3-5}
\cline{7-9}
\noalign{\vskip 0.55mm}
\parbox[t]{2mm}{\multirow{7}{*}{\rotatebox[origin=c]{90}{secondary}}} &
\multicolumn{1}{c}{type} &
\multicolumn{1}{c}{BH} &
\multicolumn{1}{c}{-} &
\multicolumn{1}{c}{BH} &
\multicolumn{1}{c}{} &
\multicolumn{1}{c}{Be star} &
\multicolumn{1}{c}{Be star} &
\multicolumn{1}{c}{Be star}
\\
\noalign{\vskip 0.55mm}
\multicolumn{1}{c}{} &
\multicolumn{1}{c}{$T_{\rm eff}^{(2)} \,[\kK]$} &
\multicolumn{1}{c}{-} &
\multicolumn{1}{c}{-} &  
\multicolumn{1}{c}{-} &
\multicolumn{1}{c}{} &
\multicolumn{1}{c}{18} &
\multicolumn{1}{c}{$18$} &
\multicolumn{1}{c}{$\in \left[18.7,19.1\right]$}
\\
\noalign{\vskip 0.55mm}
\multicolumn{1}{c}{} &
\multicolumn{1}{c}{$\Rstar^{(2)} \,[\rsun]$} &
\multicolumn{1}{c}{1} &  
\multicolumn{1}{c}{-} &
\multicolumn{1}{c}{-} &
\multicolumn{1}{c}{} &
\multicolumn{1}{c}{3.7} &
\multicolumn{1}{c}{$3.7$} &
\multicolumn{1}{c}{$\in \left[2.8,3.4\right]$}
\\
\multicolumn{1}{c}{} &  
\multicolumn{1}{c}{$\Mstar^{(2)} \,[\msun]$} &
\multicolumn{1}{c}{30} &
\multicolumn{1}{c}{-} &
\multicolumn{1}{c}{$\in \left[13,30\right]$} &
\multicolumn{1}{c}{} &
\multicolumn{1}{c}{7} &
\multicolumn{1}{c}{$\in \left[5,9\right]$} &
\multicolumn{1}{c}{$\in \left[1.6,6.9\right]$}
\\
\noalign{\vskip 0.55mm}
\multicolumn{1}{c}{} &  
\multicolumn{1}{c}{$\logg^{(2)}$} &
\multicolumn{1}{c}{-} &
\multicolumn{1}{c}{-} &  
\multicolumn{1}{c}{-} &
\multicolumn{1}{c}{} &
\multicolumn{1}{c}{4.15} &
\multicolumn{1}{c}{$4.0$} &
\multicolumn{1}{c}{$\in \left[3.7,4.3\right]$}
\\
\noalign{\vskip 0.55mm}
\multicolumn{1}{c}{} &  
\multicolumn{1}{c}{$\vrot^{(2)}\,[\kms]$} &
\multicolumn{1}{c}{-} &
\multicolumn{1}{c}{-} &
\multicolumn{1}{c}{-} &
\multicolumn{1}{c}{} &
\multicolumn{1}{c}{600} &
\multicolumn{1}{c}{$\in [400,560]$} &
\multicolumn{1}{c}{$480$}
\\
\noalign{\vskip 0.55mm}
\multicolumn{1}{c}{} &  
\multicolumn{1}{c}{$\yhe^{(2)}$} &
\multicolumn{1}{c}{-} &
\multicolumn{1}{c}{-} &
\multicolumn{1}{c}{-} &
\multicolumn{1}{c}{} &
\multicolumn{1}{c}{0.1} &
\multicolumn{1}{c}{$0.08$} &
\multicolumn{1}{c}{$0.1$}
\\
\noalign{\vskip 0.55mm}
\hline
\end{tabular}
\tablefoot{The values for the rotational velocities have
    been deduced from the inclination used in this paper (see
    Table~\ref{tab:params_lb1_orbit}, $i=22^\circ$ and $i=39^\circ$
    for the B+BH and strB+Be scenarios, respectively). For
    consistency, we calculated \logg~from the stellar radii and masses.}
\tablebib{(1) \cite{SimonDiaz2020}; (2) \cite{Lennon2021}; (3) \cite{Shenar2020}.}
\end{center}
\end{table*}
\begin{table*}
\setlength{\tabcolsep}{1.28mm}
\begin{center}
  \caption{Orbital parameters for the LB-1 system as calculated from our models and found in the literature.}
\label{tab:params_lb1_orbit}
%
%
\begin{tabular}{cccccccc}
\hline
\hline
\noalign{\vskip 0.55mm}
\multicolumn{1}{c}{} &
\multicolumn{3}{c}{models} &
\multicolumn{1}{c}{} &
\multicolumn{3}{c}{literature}
\\
\cline{2-4}
\cline{6-8}
\noalign{\vskip 0.55mm}
\multicolumn{1}{c}{} &
\multicolumn{1}{c}{PB\_SBH} &
\multicolumn{1}{c}{PB\_SBE01} &
\multicolumn{1}{c}{PB\_SBE02} &
\multicolumn{1}{c}{} &
\multicolumn{1}{c}{(1)} &
\multicolumn{1}{c}{(2)} &
\multicolumn{1}{c}{(3)}
\\
\noalign{\vskip 0.55mm}
\multicolumn{1}{c}{$i\,[^\circ]$} &
\multicolumn{1}{c}{22} &
\multicolumn{1}{c}{39} &
\multicolumn{1}{c}{39} &
\multicolumn{1}{c}{} &
\multicolumn{1}{c}{-} &
\multicolumn{1}{c}{-} &
\multicolumn{1}{c}{$\in [35,43]$}
\\
\noalign{\vskip 0.55mm}
\multicolumn{1}{c}{$K_1 \,[\kms]$} &
\multicolumn{1}{c}{53} &
\multicolumn{1}{c}{53} &
\multicolumn{1}{c}{53} &
\multicolumn{1}{c}{} &
\multicolumn{1}{c}{53} &
\multicolumn{1}{c}{53} &
\multicolumn{1}{c}{53}
\\
\noalign{\vskip 0.55mm}
\multicolumn{1}{c}{$K_2^{\rm (true)} \,[\kms]$} &
\multicolumn{1}{c}{7.4} &
\multicolumn{1}{c}{9.9} &
\multicolumn{1}{c}{9.9} &
\multicolumn{1}{c}{} &
\multicolumn{1}{c}{-} &
\multicolumn{1}{c}{$\in [8,13]$} &
\multicolumn{1}{c}{$\in [10.2,12.2]$}
\\
\noalign{\vskip 0.55mm}
\multicolumn{1}{c}{$K_2^{\rm (H_\alpha)} \,[\kms]$} &
\multicolumn{1}{c}{10} &
\multicolumn{1}{c}{13} &
\multicolumn{1}{c}{8} &
\multicolumn{1}{c}{} &
\multicolumn{1}{c}{$\in [5.6,7.2]$} &
\multicolumn{1}{c}{-} &
\multicolumn{1}{c}{-}
\\
\noalign{\vskip 0.55mm}
\hline
\end{tabular}
%
%
\tablebib{(1) \cite{Liu2019}; (2) \cite{Liu2020}; (3) \cite{Shenar2020}.}
\end{center}
\end{table*}
\begin{table}
\setlength{\tabcolsep}{0.78mm}
\begin{center}
\caption{Disc parameters for the three different (best-fit) models as
    obtained from our simulations of the \mbox{LB-1} system (see text).}
\label{tab:params_lb1_fit}
%
%
\begin{tabular}{ccccc}
\hline
\hline
\noalign{\vskip 0.55mm}
\multicolumn{1}{c}{} &
\multicolumn{1}{c}{model} &
\multicolumn{1}{c}{PB\_SBH} &
\multicolumn{1}{c}{PST\_SBE01} &
\multicolumn{1}{c}{PST\_SBE02}
\\\hline
\noalign{\vskip 0.55mm}
\parbox[t]{2mm}{\multirow{6}{*}{\rotatebox[origin=c]{90}{primary}}} &
\multicolumn{1}{c}{type} &
\multicolumn{1}{c}{B star} &
\multicolumn{1}{c}{stripped star} &
\multicolumn{1}{c}{stripped star}
\\
\noalign{\vskip 0.55mm}
\multicolumn{1}{c}{} &  
\multicolumn{1}{c}{$\vmicro^{(1)} \,[\kms]$} &
\multicolumn{1}{c}{-} &
\multicolumn{1}{c}{-} &
\multicolumn{1}{c}{$30$}
\\
\noalign{\vskip 0.55mm}
\multicolumn{1}{c}{} &  
\multicolumn{1}{c}{$\rho_0^{\rm (1)}\,[\gcmc]$} &
\multicolumn{1}{c}{-} &
\multicolumn{1}{c}{-} &
\multicolumn{1}{c}{$1.3 \cdot 10^{-12}$}
\\
\noalign{\vskip 0.55mm}
\multicolumn{1}{c}{} &  
\multicolumn{1}{c}{$\beta_{\rm D}$} &
\multicolumn{1}{c}{-} &
\multicolumn{1}{c}{-} &
\multicolumn{1}{c}{$7/2$}
\\
\noalign{\vskip 0.55mm}
\multicolumn{1}{c}{} &  
\multicolumn{1}{c}{$T_{\rm disc}\,[\kK]$} &
\multicolumn{1}{c}{-} &
\multicolumn{1}{c}{-} &
\multicolumn{1}{c}{Eq.~\eqref{eq:tdisc}}
\\\hline
\noalign{\vskip 0.55mm}
\parbox[t]{2mm}{\multirow{6}{*}{\rotatebox[origin=c]{90}{secondary}}} &
\multicolumn{1}{c}{type} &
\multicolumn{1}{c}{BH} &
\multicolumn{1}{c}{Be star} &
\multicolumn{1}{c}{Be star}
\\
\noalign{\vskip 0.55mm}
\multicolumn{1}{c}{} &  
\multicolumn{1}{c}{$\vmicro^{(2)} \,[\kms]$} &
\multicolumn{1}{c}{$100$} &
\multicolumn{1}{c}{$100$} &
\multicolumn{1}{c}{$100$}
\\
\noalign{\vskip 0.55mm}
\multicolumn{1}{c}{} &  
\multicolumn{1}{c}{$\rho_0^{\rm (2)}\,[\gcmc]$} &
\multicolumn{1}{c}{$3\cdot 10^{-12}$} &
\multicolumn{1}{c}{$6.9\cdot 10^{-13}$} &
\multicolumn{1}{c}{$6.4\cdot 10^{-13}$}
\\
\noalign{\vskip 0.55mm}
\multicolumn{1}{c}{} &  
\multicolumn{1}{c}{$\beta_{\rm D}$} &
\multicolumn{1}{c}{$3/2$} &
\multicolumn{1}{c}{$3/2$} &
\multicolumn{1}{c}{$3/2$}
\\
\noalign{\vskip 0.55mm}
\multicolumn{1}{c}{} &  
\multicolumn{1}{c}{$T_{\rm disc}\,[\kK]$} &
\multicolumn{1}{c}{$=\const=8$} &
\multicolumn{1}{c}{Eq.~\eqref{eq:tdisc}} &
\multicolumn{1}{c}{Eq.~\eqref{eq:tdisc}}
\\
\noalign{\vskip 0.55mm}
\hline
\end{tabular}
%
\end{center}
\end{table}
In the following, we describe our model for the B+BH scenario (model
PB\_SBH, primary: B star, secondary: BH).  For the B star, we used the
stellar parameters as summarized in Table~\ref{tab:params_lb1}. With
$K_1$ and $K_2^{\rm (true)}$ given, the mass of the black hole is
constrained via Eqs.~\eqref{eq:orbit_binary1} and
\eqref{eq:orbit_binary2} yielding $M_2 = M_1 K_1/K_2^{\rm (true)}
\approx [20,30]\,\msun$, where we have chosen the black-hole mass as
$M_2=30\,\msun$ and inclination $i=22^\circ$ in order to reproduce the
semi-amplitude of the radial velocity $K_1$. With this inclination, we
adopted a rotational velocity for the B star, $\vrot^{(1)}=21\,\kms$
(with the observed value $\vsini = 8 \,\kms$,
\citealt{Lennon2021}). To model the disc, we have set the minimum radius to
$\Rstar^{(2)} = 1\,\rsun$, since any regions with smaller radii would
not contribute to the emission due to the small emitting area in any
case, and have set the maximum radius of the disc to the distance from the
BH to the first Lagrange point.

For a standard $\alpha$-disc model without irradiation from an
external source, we do not expect any emission in \Ha, since the
surface of the hot regions close to the BH is very small, and the
B-star companion completely dominates the total radiation flux. In the
B+BH binary system, however, the disc will also be heated by the
B-star companion. In order to avoid complex multi-D
radiation-hydrodynamic simulations, and to keep our model as simple as
possible, we introduced a constant temperature within the disc as a
free parameter. With given stellar parameters and inclination then,
the model is completely specified by four input parameters, namely the
disc temperature, $T_{\rm disc}$, the micro-turbulent velocity in the
disc, $\vmicro^{(2)}$, the base density of the disc, $\rho_0^{(2)}$,
and the slope of the density stratification, $\beta_{\rm D}^{(2)}$.

We then calculated \Ha~line profiles with the algorithm described in
Sect.~\ref{sec:binary_theory} for different models (see
Appendix~\ref{app:parameter_study_pbsbh}). As an inner boundary
condition for the specific intensity emerging the B star, we used a
photospheric line profile obtained from \textsc{FASTWIND}. Further, we
calculated the radial velocity curve from the barycentric method
considering only the wings of the \Ha~lines up to 1/3 of the total
height of the \Ha~line profile as suggested by \cite{Liu2019}.  The
results for our best fit (by eye) with disc temperature, $T_{\rm
  disc}=8\,\kK$, micro-turbulent velocity $\vmicro^{\rm
  (2)}=100\,\kms$, base density $\rho_0^{(2)}=3\cdot 10^{-12}\,\gcmc$,
and radial density stratification $\beta_{\rm D}=3/2$ are shown in
Fig.~\ref{fig:profiles_best}, with corresponding density,
  temperature, and velocity profiles in the equatorial plane shown in
  Fig.~\ref{fig:models_best}.

The slope of the density stratification is in fairly good agreement
with accretion discs described within the $\alpha$-disc
prescription. Indeed, a shallow slope of the density stratification is
also required from a modelling point of view to obtain a centrally
peaked emission profile. For a steeper decline of the density, for
instance, the outer disc regions (where the velocities are lowest)
barely contribute to the low-velocity emission due to much smaller
densities.

On the other hand, the emitting area at high (projected) velocities (\ie~near to
the inner disc radius) is quite small. Thus, a relatively high
micro-turbulent velocity is required to obtain the broad \Ha~emission
wings, effectively increasing the emitting area (and consequently the
emerging flux) at high velocity shifts. In particular, this also
smoothens the typically double-peaked emission feature of accretion
discs even at such inclinations ($i=22^\circ$). High turbulent
velocities, however, are frequently found when modelling dynamical
systems with simplified stationary models (which by assumption neglect
small-scale turbulent motions in the gas).  As just one example, to
mimick the observed velocity dispersion when analysing magnetically
confined winds described with a steady state model, \cite{Owocki16}
convolved their synthetic \Ha~line profiles with a $150\,\kms$
Gaussian `macro-turbulence' velocity. Since we have neglected
gravitational interactions of the primary object with the disc, we
might thus expect the inferred high turbulent velocities to mimick
dynamical effects within the binary system.

The qualitative behaviour of the observed \Ha~line profile is well
reproduced, with corresponding phase-averaged equivalent width
measured in velocity space in units of $100\,\kms$, $\bar{W}_x= 6.81$,
in good agreement with observations ($\bar{W}_x^{\rm (obs)}=6.84$). We
emphasize, however, that the radial velocity curves show some
discrepancies when compared to observations (see also
Table~\ref{tab:params_lb1_orbit}). While the actual orbital
velocities, $K_2^{(\rm true)} = v_2\sin\left(i\right)=7.4\,\kms$, are
slightly below the observed range ($K_2^{(\rm true, obs)} \in
[8,13]\,\kms$), we find a higher semi-amplitude when applying the
barycentric method ($K_2^{\rm (H_\alpha)}=10\,\kms$). Indeed, this
discrepancy can be explained by the absorption profile of the B star
(see discussion in \citealt{AbdulMasih2020}), and could be solved by
an additional \Ha~emission component attached, for example, to the B
star or to a tertiary object. Such an additional emission component
would reduce the radial velocity amplitude as measured from the
barycentric method, by slightly modifying the \Ha~line wings.

More pronounced, we find significant differences between the observed
and synthetic (mean-subtracted) dynamical spectra of the
\Ha~line. While the observed dynamical spectrum clearly displays an
anti-phase behaviour of the line wings and the line core, the
corresponding features in the synthetic dynamical spectrum are fairly
well aligned. As shown below for the strB+Be scenario, this issue
could be solved by introducing also a disc around the B star though.

A main theme of our model regards the temperature stratification of
the BH disc, where we required $T_{\rm disc}=8\,\kK$ in order to
reproduce the observations. This temperature region can indeed be
considered as a sweet spot, since lower temperatures ($T_{\rm
  disc}\lesssim 5\,\kK$) would require much higher densities to
provide enough emission in \Ha. The corresponding densities would
translate to a high mass-accretion rate of the BH within the standard
$\alpha$-disc prescription, thus predicting an X-ray bright accretion
disc that has not been observed. Similarly, too high
temperatures ($T_{\rm disc}\gtrsim 15\,\kK$) are not allowed either,
because we would need to increase the densities to match the
equivalent width of the observed \Ha~line profiles (see
  Appendix \ref{app:parameter_study_pbsbh}), again yielding an
  X-ray bright accretion disc. If the sweet-spot temperature as found
in our model is a reasonable substitute for detailed numerical
simulations still needs to be investigated.
\subsection{Stripped-star and Be-star (strB+Be) scenario}
\label{subsec:lb1a}
For the stellar parameters of the \mbox{LB-1} system within the
strB+Be scenario, we adopted the values proposed by \cite{Shenar2020},
summarized again in Table~\ref{tab:params_lb1}, and yielding an
inclination $i=39^\circ$. Motivated by these authors, we have set the
rotational velocity of the stripped star to $v_{\rm rot}^{\rm (1)}=11$
(with $\vsini = 7\,\kms$ following \citealt{Shenar2020} and
\citealt{Lennon2021}), and assumed the Be-star to rotate at its
critical velocity, $v_{\rm rot}^{\rm (2)}=\sqrt{G \Mstar^{\rm (1)} /
  \Rstar^{\rm(1)}}$. For simplicity, we neglected effects of gravity
darkening.

To set the temperature stratification in the disc, we followed the
qualitative behaviour as found by detailed radiative-transfer
calculations in \citet[][Sect.~5.2]{Carciofi2006}. In regions close to
the stellar surface (where the vertical electron-scattering optical
depth $\tau_{z}(r \sin\clatitude)>0.1$), we adopted an analytical
temperature stratification described by a flat blackbody reprocessing
disc (their Eq.~(12)):
\beq
\label{eq:tdisc}
T_{\rm disc} \left( r,\clatitude \right) = \dfrac{T_0}{\pi^{1/4}} \left[\sin^{-1}\left( \dfrac{\Rstar}{r \sin \clatitude}\right)-\dfrac{\Rstar}{r \sin \clatitude}\sqrt{1-\dfrac{\Rstar^2}{r^2\sin^2\clatitude}} \right]^{1/4} \,,
\eeq
with $T_0$ the base temperature of the disc. Since the base densities
of our disc models are typically relatively low, we can neglect
back-warming effects of the photosphere by the disc, and therefore have set
the base temperature to the stellar effective temperature,
$T_0=\Teff$. To mimick the heating of the disc in (vertically)
optically thin regions by the star, we adopted a constant disc
temperature $T_{\rm disc}=0.6\cdot\Teff$ where $\tau_{z}(r
\sin\clatitude)<0.1$ (again, following the argumentation by
\citealt{Carciofi2006}).

We note already here that we additionally allowed for a disc attached
also to the stripped star (see below). With given stellar parameters,
a now specified temperature stratification, and a given inclination to
reproduce the semi-amplitude of the radial velocity curve of the
primary, $K_1$, the model is completely specified by six input
parameters, namely the micro-turbulent velocities, $\vmicro^{\rm
  (1,2)}$, the base density $\rho_0^{\rm (1,2)}$, and the slope of the
density stratification of both discs, $\beta_{\rm D}^{\rm
  (1,2)}$. When considering only the Be-star disc, the number of free
parameters is reduced to three.

As for the B+BH scenario above, we calculated synthetic \Ha~line
profiles with the developed \textsc{BOSS-3D} code for various
different models (see Appendix~\ref{app:parameter_study_pstrsbe02}),
with the photospheric line profiles of both stars again obtained from
\textsc{FASTWIND}. The results for our best two models are shown in
Fig.~\ref{fig:profiles_best}, with the corresponding
  density, temperature, and velocity stratification in the equatorial
  plane shown in Fig.~\ref{fig:models_best}, and the orbital and disc
  parameters summarized in Tables~\ref{tab:params_lb1_orbit} and \ref{tab:params_lb1_fit}, respectively.
\paragraph{Model PST\_SBE01.}
This model describes a stripped star (primary object) in orbit with a
Be-star (secondary object) hosting a circumstellar disc. The obtained
disc parameters, $\beta_{\rm D}^{\rm (2)}=3/2$ and $\rho_0^{\rm
  (2)}=6.9\cdot 10^{-13}\,\gcmc$ are both at the lower end of expected
values (see above and \citealt{Silaj2010}), which can be explained as
follows: Firstly, the inner regions of the disc (\ie~those near to the
star with highest rotational velocities) would become optically thick
in the continuum\footnote{We have approximated the continuum by pure
  electron-scattering opacities and assuming LTE to determine the
  source function.} for higher base densities, and the broad line
wings would vanish within the continuum flux. Secondly, as described
above for the BH disc, a shallow slope of the density stratification
is required to obtain a centrally peaked emission profile. Again, we
also require a high micro-turbulent velocity to obtain the broad
\Ha~emission wings and to smooth out the otherwise double-peaked
emission profile.

Even for our best fit (by eye) parameters, however, the central
emission peak is not well reproduced with this model. Additionally,
the equivalent width averaged over all phases ($\bar{W}_x=6.57$, again
measured in $100\,\kms$) is somewhat smaller than expected from
observations ($\bar{W}_x^{\rm (obs)}=6.84$). Moreover, the obtained
semi-amplitude of the radial velocity of the secondary object,
$K_2^{\rm (H_\alpha)}=13\,\kms$, shows a clear deviation from the
actually observed values ($K_2^{\rm (H_\alpha, obs)}\approx
6\,\kms$). As in the B+BH scenario, the semi-amplitude is larger than
the orbital velocity of the secondary object ($v_2\sin(i)=9.9\,\kms$),
demonstrating again the impact of the photospheric absorption profile
underlying the stripped star in this case. Finally, the dynamical
\Ha~line profile shows the same behaviour as for the B+BH scenario, in
contrast to observations.

\paragraph{Model PST\_SBE02.}
To improve the model and to increase the low-velocity emission, we
additionally implemented a circumstellar disc attached to the stripped
star. Although speculative, this disc might have formed by
re-accretion from the Be-star (\eg~\citealt{Shenar2020}) or from
earlier mass-transfer phases.

While the best fit (by eye) gives a similar disc for the Be-star as
found above, we predict a slightly higher base density of the
stripped-star disc, $\rho_0^{\rm (1)}=1.3\cdot 10^{-12}\,\gcmc$, and a
relatively steep slope of the density stratification, $\beta_{\rm
  D}^{\rm (1)}=7/2$. Since the orbital velocities of the stripped-star
disc are quite low in any case (Eq.~\eqref{eq:vrot} with a small mass
and large radius), this model gives an increased central emission
peak, at least for moderate micro-turbulent velocities. The
qualitative behaviour of the observed \Ha~line profile then is very
well reproduced, with corresponding equivalent width, $\bar{W}_x^{\rm
  (obs)}=6.84$, in very good agreement with observations. Since the
\Ha~line wings are slightly contaminated by the (anti-phase) emission
from the stripped-star disc, the obtained radial velocity curve
($K_2^{\rm (H_\alpha)}=8 \,\kms$) agrees fairly well with the
corresponding observations ($K_2^{\rm (H_\alpha, obs)}=6\,\kms$ as
derived from the observed \Ha~line wings).  This underestimation of
the true orbital velocities due to the contamination of the line
profile by an emission component from the primary object is tightly
connected to the opposite effect as found by \cite{AbdulMasih2020} for
an absorption component.

While we were not aiming at an exact reproduction of the radial
velocity curves due to the simplicity of our disc model
(\eg~neglecting disc inhomogeneities), we show that the disc's
emission component can qualitatively explain the discrepancies between
the true and apparent orbital motion of the system.  We emphasize that
any sort of relatively narrow, centrally peaked \Ha~emission
originating from the stripped star might provide similar results. The
origin of such narrow \Ha~emission, however, remains unclear. For
instance, we were not able to reproduce the observations with a simple
($\beta$-velocity-type) stellar wind blown from the stripped star. In
contrast, the dynamical spectrum of our strB+Be model with an
additional disc around the stripped star is in good agreement with
observations. Thus, and due to the simplicity of our disc model (with
only three free parameters for each disc), we propose that \mbox{LB-1}
contains two discs, each attached to the individual objects in a
strB+Be binary system (as shown here), or in a B+BH binary (as
qualitatively explained above).  We emphasize that \cite{Shenar2020}
found a centrally peaked \Ha~emission from the stripped star by using
disentangling techniques as well, suggesting that this residual
component might originate from a disc around the stripped star. Our
present results put this interpretation on a much firmer ground.
%
%
\section{Summary and conclusions} \label{sec:conclusions}
In this paper, we have presented a new method for calculating
synthetic spectra of binary systems, assuming the orbital setup and
the atmospheric and circumstellar structure(s) to be known. This
method can be extended to multiple systems, with the computation time
scaling only linearly (or slightly super-linearly) with the number of
involved objects. Moreover, by assigning individual coordinate systems
to each object and accounting for the appropriate coordinate
transformations, the developed method automatically accounts for the
different length-scales of each object. The presented algorithm and
associated code (\textsc{BOSS-3D}) is therefore capable of calculating
synthetic line profiles for a wide variety of astrophysical systems,
such as planetary or stellar transits, jets associated with young
stellar objects (YSO's) or post-asymptotic giant branch binary systems
(see also \citealt{Bollen2020}), and multiple systems involving
(possibly colliding) stellar winds, Be stars, or black holes (BH's)
with discs.

As a first application of the method, we considered the \Ha~line
formation of the B-star/BH-disc (B+BH) and stripped-star/Be-star
(strB+Be) scenarios for the enigmatic (multiple) system \mbox{LB-1}
that have previously been proposed (among other hypotheses, see
Sect.~\ref{sec:intro}) by \cite{Liu2019} and \cite{Shenar2020},
respectively. To calculate the density stratification and velocity
field, we applied an analytical, geometrically thin, Keplerian disc
model in vertical hydrostatic equilibrium. Further, we assigned a
constant temperature to the BH disc as a free parameter, and used a
prescribed temperature stratification based on \cite{Carciofi2006} for
the Be-star disc. The level populations for the \Ha~line transition
have then been computed from Saha-Boltzmann statistics assuming full
ionization.

Under these assumptions none of the models can reproduce the detailed
phase-dependent shape of the observed line profiles, particularly
due to differences in the dynamical spectrum (and missing
  low-velocity emission in some cases).  Moreover, the radial
velocity curves measured from the synthetic \Ha~line wings are highly
overestimated due to the absorption component of the primary object
(see also \citealt{AbdulMasih2020}). It remains open whether
  NLTE calculations can help to explain the detailed phase-dependent
  shape of \Ha~line profiles. Detailed theoretical investigations are
  needed to test these effects. Alternatively, we have empirically set
  an additional disc around the stripped star within the strB+Be
  scenario to solve this problem, and found a sound reproduction of
  the observed \Ha~line profiles both from their qualitative shape and
  their equivalent widths, as well as of the dynamical spectrum and
  the radial velocity curves. This putative disc might be a remnant
  associated with earlier mass-transfer phases or could have
  originated from re-accretion of material from the Be-star disc.

Similarly, such a disc might be attached to the B star in the B+BH
scenario as well, and the B+BH hypothesis remains a valid possibility
to explain the \mbox{LB-1} system.  In any case, our findings provide
strong evidence that \mbox{LB-1} contains a disc-disc system, with an
additional disc either attached to the B star (in the B+BH scenario)
or to the stripped star (in the strB+Be scenario).
\begin{acknowledgements}
We thank our referee, Dr. Maria Bergemann, for many helpful comments
and suggestions.  LH and JOS gratefully acknowledge support from the
Odysseus program of the Belgian Research Foundation Flanders (FWO)
under grant G0H9218N.  JB acknowledges support from the FWO Odysseus
program under project G0F8H6N.  The National Solar Observatory (NSO)
is operated by the Association of Universities for Research in
Astronomy, Inc. (AURA), under cooperative agreement with the National
Science Foundation.  TS acknowledges funding received from the
European Research Council (ERC) under the European Union's Horizon
2020 research and innovation programme (grant agreement number 772225:
MULTIPLES) and from the European Union’s Horizon 2020 under the Marie
Sk\l odowska-Curie grant agreement No 101024605. IEM has received
funding from the European Research Council (ERC) under the European
Union’s Horizon 2020 research and innovation programme (SPAWN ERC,
grant agreement No 863412).
\end{acknowledgements}
%
%
\bibliographystyle{aa}
\bibliography{bib_levin}
%
%
\appendix
%
%
\section{Coordinate transformations}
\label{app:transformations}
\begin{figure}[t]
\resizebox{\hsize}{!}{\includegraphics{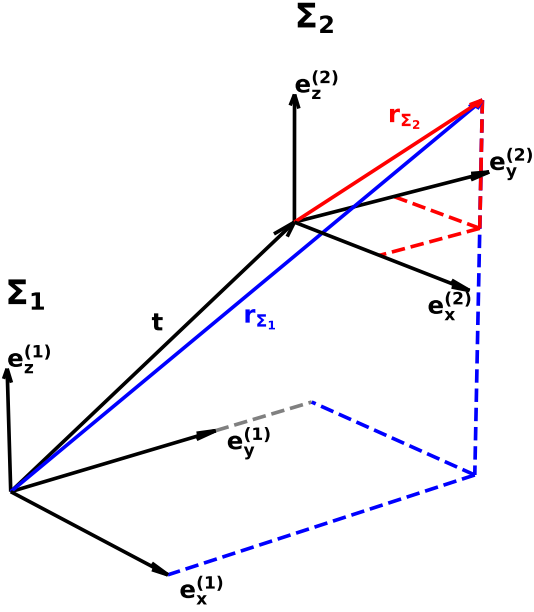}}
\caption{Transformation between coordinate systems
  $\Sigma_1=\left(\vecown{e}_x^{(1)},\vecown{e}_y^{(1)},\vecown{e}_z^{(1)}
  \right)$ and $\Sigma_2 =
  \left(\vecown{e}_x^{(2)},\vecown{e}_y^{(2)},\vecown{e}_z^{(2)}
  \right)$, where $\vecown{t}$ indicates the translation vector from
  $\Sigma_1\rightarrow \Sigma_2$, and $\vecown{r}_{\Sigma_1}$ and
  $\vecown{r}_{\Sigma_2}$ denote the coordinate representations in
  each system for a given point.}
%
\label{fig:app_transformation}
\end{figure}

In this section, we derive the basic transformations of coordinates
between two coordinate systems (see Fig.~\ref{fig:app_transformation})
with orthogonal basis such that
$\Sigma_1=\left(\vecown{e}_x^{(1)},\vecown{e}_y^{(1)},\vecown{e}_z^{(1)}
\right)$ and
$\Sigma_2=\left(\vecown{e}_x^{(2)},\vecown{e}_y^{(2)},\vecown{e}_z^{(2)}
\right)$. Any point $\vecown{r}_{\Sigma_1}$ in $\Sigma_1$ can then be
expressed by
\beqa
\nonumber
\vecown{r}_{\Sigma_1} &=& x_1 \vecown{e}_x^{(1)} + y_1 \vecown{e}_y^{(1)} + z_1 \vecown{e}_z^{(1)} \\
\label{eq:app_trans0}
&=& s_x x_2 \vecown{e}_x^{(2)} + s_y y_2 \vecown{e}_y^{(2)} + s_z z_2 \vecown{e}_z^{(2)} + \vecown{t} \,,
\eeqa
where $(x_1,y_1,z_1)$ and $(x_2,y_2,z_2)$ are the coordinates in each
system, $s_{x,y,z}$ are scale factors allowing us to include different
length scales for both coordinate systems, $\vecown{t}=t_x
\vecown{e}_x^{(1)} + t_y \vecown{e}_y^{(1)} + t_z \vecown{e}_z^{(1)}$
is the translation vector indicating the origin of $\Sigma_2$ within
$\Sigma_1$, and $\vecown{e}_{x,y,z}^{(2)}$ are the basis vectors of
$\Sigma_2$ measured within $\Sigma_1$.

For our purposes, the length scales of both coordinate systems are
constant (and typically refer to the radius of each individual
object).  Thus, we define $s:= L_{\Sigma_2} /
L_{\Sigma_1}=s_x=s_y=s_z$, with $L_{\Sigma_2}$ and $L_{\Sigma_1}$
describing the length scales of the individual coordinate
systems. Multiplying Eq.~\eqref{eq:app_trans0} with
$\vecown{e}_x^{(1)}$, $\vecown{e}_y^{(1)}$, and $\vecown{e}_z^{(1)}$,
we obtain in matrix form:
\beqa
\nonumber
\underbrace{
\begin{pmatrix}
x_{1} \\
y_{1} \\
z_{1} 
\end{pmatrix}
}_{\vecown{r}_{\Sigma_1}}
&=&
\underbrace{
\begin{pmatrix}
s & 0 & 0 \\
0 & s & 0 \\
0 & 0 & s
\end{pmatrix}
}_{=:\matown{S}}
\underbrace{
\begin{pmatrix}
\vecown{e}_x^{(2)} \cdot \vecown{e}_x^{(1)} & \vecown{e}_y^{(2)} \cdot \vecown{e}_x^{(1)} & \vecown{e}_z^{(2)} \cdot \vecown{e}_x^{(1)} \\
\vecown{e}_x^{(2)} \cdot \vecown{e}_y^{(1)} & \vecown{e}_y^{(2)} \cdot \vecown{e}_y^{(1)} & \vecown{e}_z^{(2)} \cdot \vecown{e}_y^{(1)} \\
\vecown{e}_x^{(2)} \cdot \vecown{e}_z^{(1)} & \vecown{e}_y^{(2)} \cdot \vecown{e}_z^{(1)} & \vecown{e}_z^{(2)} \cdot \vecown{e}_z^{(1)}
\end{pmatrix}
}_{=:\matown{A}}
\underbrace{
\begin{pmatrix}
x_{2} \\
y_{2} \\
z_{2} 
\end{pmatrix}
}_{\vecown{r}_{\Sigma_2}}
\\
&+&
\underbrace{
\begin{pmatrix}
t_{x} \\
t_{y} \\
t_{z} 
\end{pmatrix}
}_{\vecown{t}}
\,.
\eeqa
Thus, the transformations between the coordinate systems $\Sigma_1
\leftrightarrow \Sigma_2$ are readily given by:
\beqa
\vecown{r}_{\Sigma_1} &=& \left[\matown{S}\matown{A}\right] \cdot \vecown{r}_{\Sigma_2} + \vecown{t} \\
\vecown{r}_{\Sigma_2} &=& \left[\matown{S}\matown{A}\right]^{-1} \cdot \left[ \vecown{r}_{\Sigma_1} - \vecown{t} \right] \,.
\eeqa
%
%
\section{Test calculations}
\label{app:tests}
\begin{figure}[t]
 \resizebox{\hsize}{!}{\includegraphics{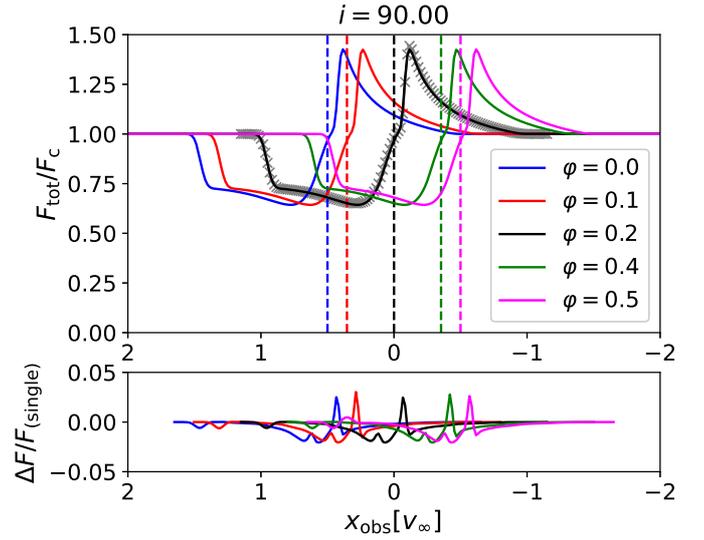}}
 \caption{Top panel: Synthetic line profiles as a function of
   frequency shift from line centre in units of the terminal velocity
   for our spherically symmetric test model (see text). The solid
   lines indicate the line profiles as obtained from the binary code
   at different phases $\varphi$ and a fixed inclination $i=90^\circ$,
   and the grey crosses correspond to the solution as obtained from
   the single-star code. The vertical dashed lines indicate the
   transition frequency of the line at the different phases for the
   assumed orbital configuration. Bottom panel: Relative error of the
   obtained line profiles at different phases when compared to the
   single-star code (shifted by the corresponding projected orbital
   velocities).}
%
\label{fig:fluxem_test00}
\end{figure}
In this section, we test the developed \textsc{BOSS-3D} code by
calculating synthetic line profiles of a generic resonance-line
transition within a spherically symmetric wind of a prototypical hot,
massive star. In order to compare the results with the single-object
code described in Sect.~\ref{sec:single_theory}, the secondary object
is assumed to be dark ($\Teff^{(2)}=10\,\kelvin$) and small
($\Rstar^{(2)}=0.2\,\rsun$). The coordinate system of the secondary
object, however, has been set to a large extent, in order to test if
the routines for setting up the triangulation and the individual rays
are performing reasonably well.  By assigning an artificial circular
orbit with absolute velocity $v_{\rm orb} = 1000 \,\kms$ to the
primary object, we further test if the corresponding Doppler-shifts
are correctly implemented for different viewing angles. The primary
object is described by $\Rstar^{(1)}=20 \,\rsun$, $\Teff^{(1)} = 40
\,\kK$, and with a wind stratification given from a $\beta$-velocity
law with base velocity $\vmin=10\,\kms$, terminal velocity $\vinf=2000
\,\kms$, and $\beta=1$. The required density is obtained from the
continuity equation assuming a mass-loss rate $\mdot=10^{-6} \msunyr$,
and the source function is calculated from the Sobolev approximation
(\citealt{Sobo60}) for a pure scattering line within the two-level
approach. Finally, we specify the boundary condition by the Planck
function, $I_\nu = B_\nu(\Teff^{(1,2)})$ , if a ray hits the core of
one of the objects.

Figure~\ref{fig:fluxem_test00} displays the resulting line profiles for
edge-on observer directions at different phases, compared with the
result from the single-object code. Since the relative error remains
below few percent, we are highly confident that the algorithm also
works for more complex simulations. Moreover, the expected velocity
shift at the different phases is exactly reproduced. We finally
emphasize that we obtain the same results when interchanging the
primary and secondary object within our algorithm, as required.
%
%
\section{Parameter study}
\label{app:parameter_study}
In this section, we perform a parameter study of the disc model
introduced in Sect.~\ref{sec:lb1}, and investigate the main effects of
the different model parameters on the various measured quantities
(\ie~the line-profile shape, the equivalent width, the radial-velocity
curve and its semi-amplitude). To this end, we focus on the
B-star and BH-disc scenario (model PB\_SBH), and on the
stripped-star and Be-star scenario with two discs (model PST\_SBE02). The
individual disc parameters are varied one at a time, starting from our
best-fit by eye models presented in Sect.~\ref{sec:lb1} and summarized
in Table~\ref{tab:params_lb1_fit}.  The radial-velocity curves have
been derived from the \Ha~line wings defined as those parts of the
synthetic line profile with normalized flux below a threshold of $1/3$
of the flux peak (see also \citealt{Liu2019}).
\subsection{Parameter study for the B-star and BH-disc scenario}
\label{app:parameter_study_pbsbh}
\begin{figure*}[t]
  \centering
  \resizebox{0.33\hsize}{!}{\includegraphics{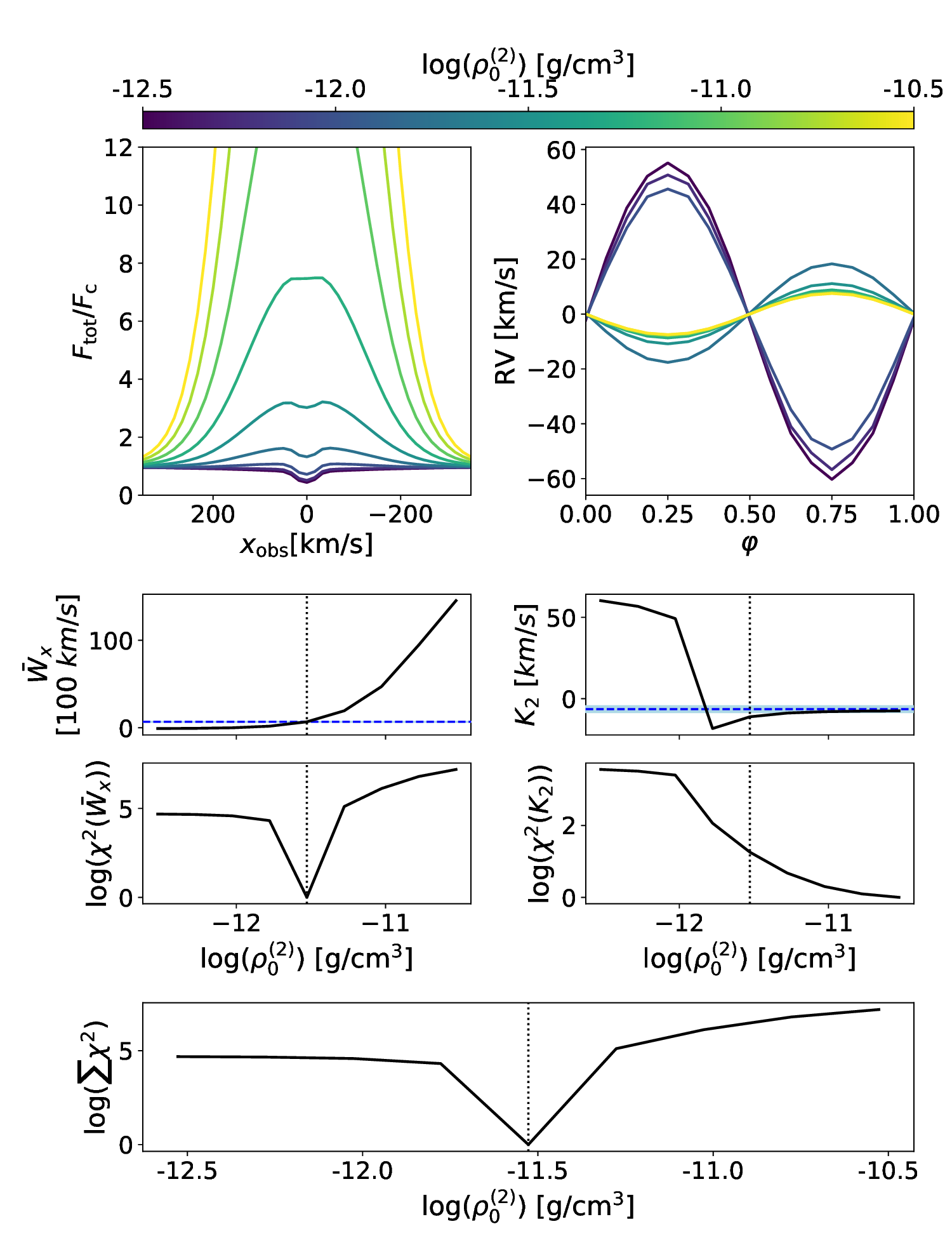}}
  \resizebox{0.33\hsize}{!}{\includegraphics{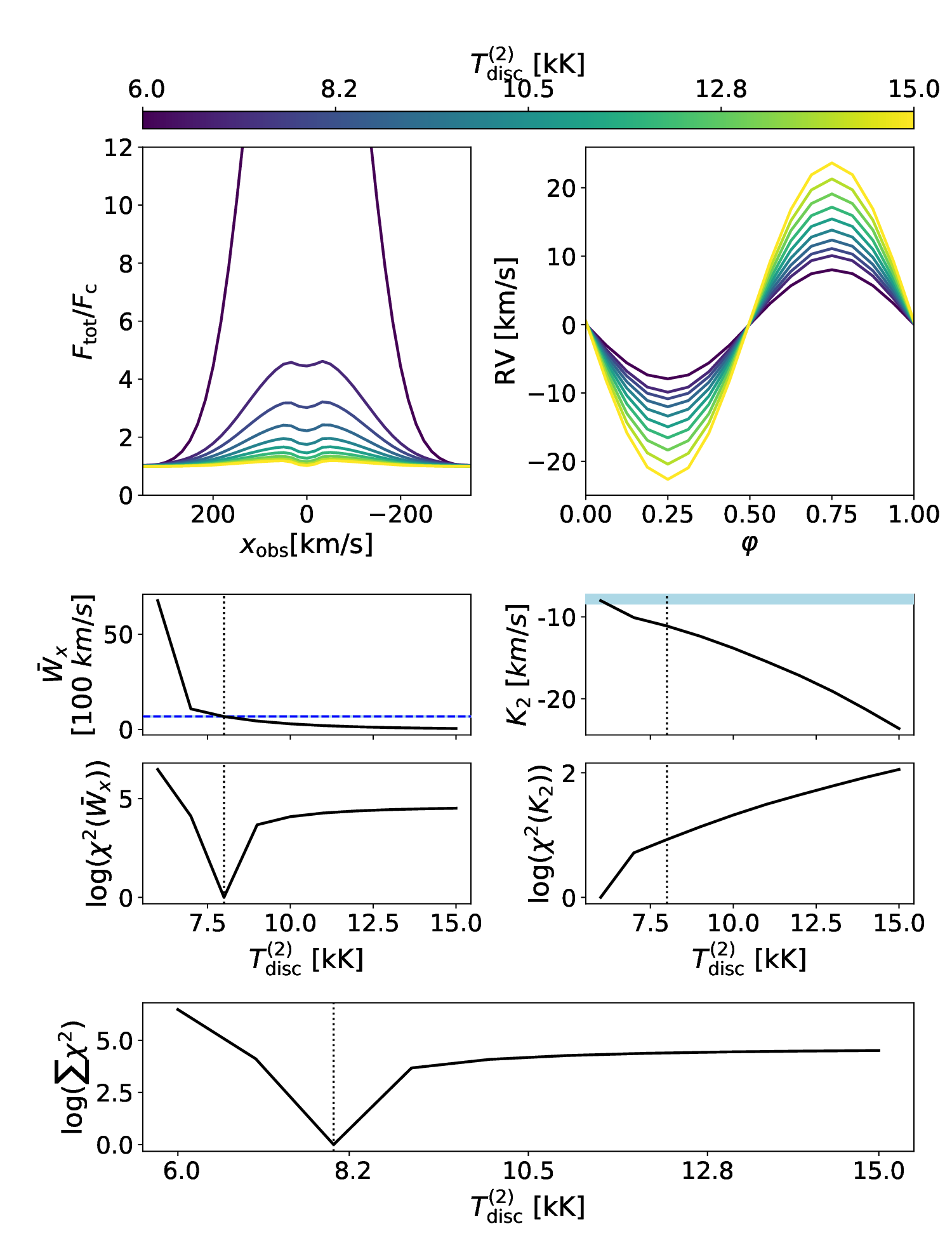}}\\
  \resizebox{0.33\hsize}{!}{\includegraphics{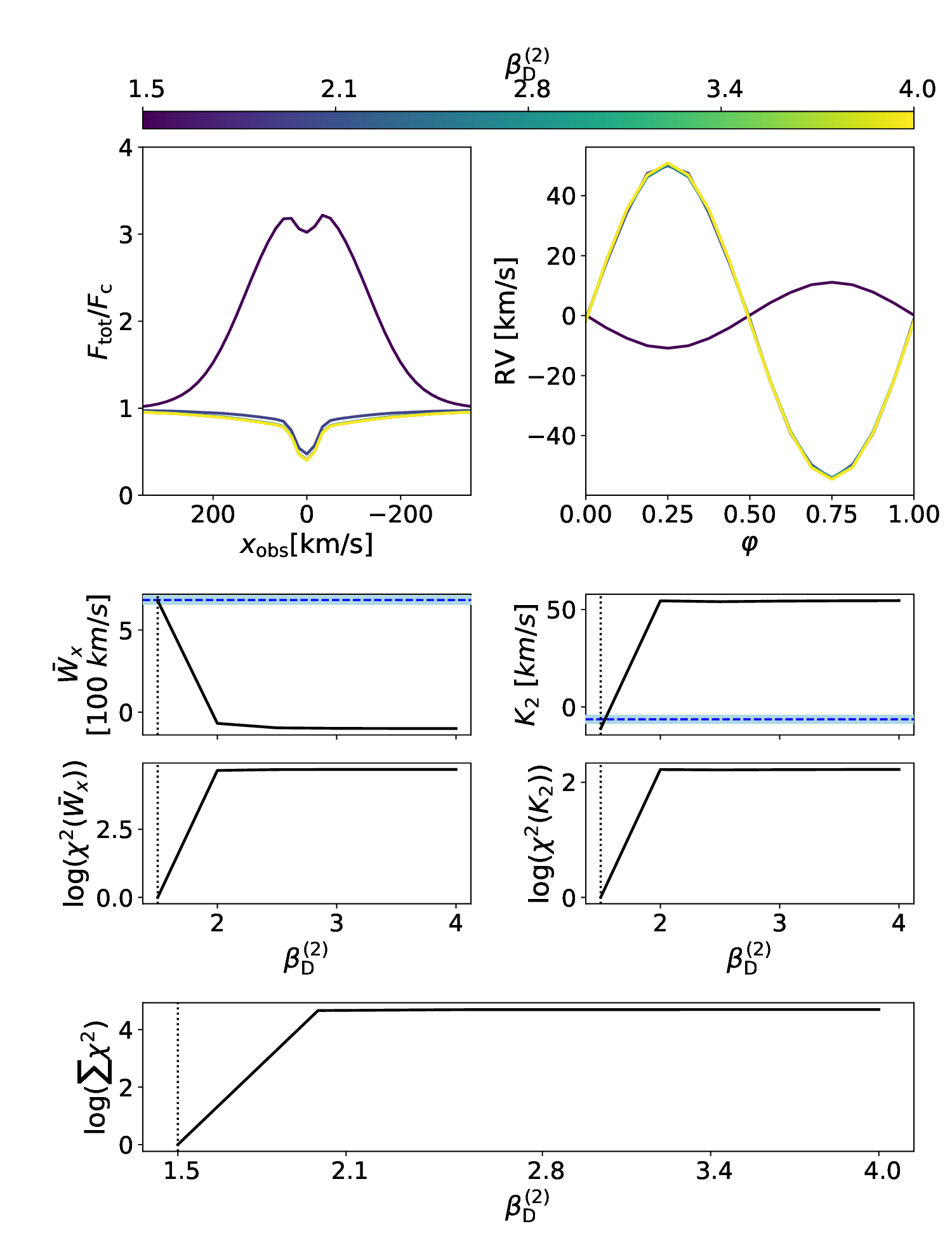}}
  \resizebox{0.33\hsize}{!}{\includegraphics{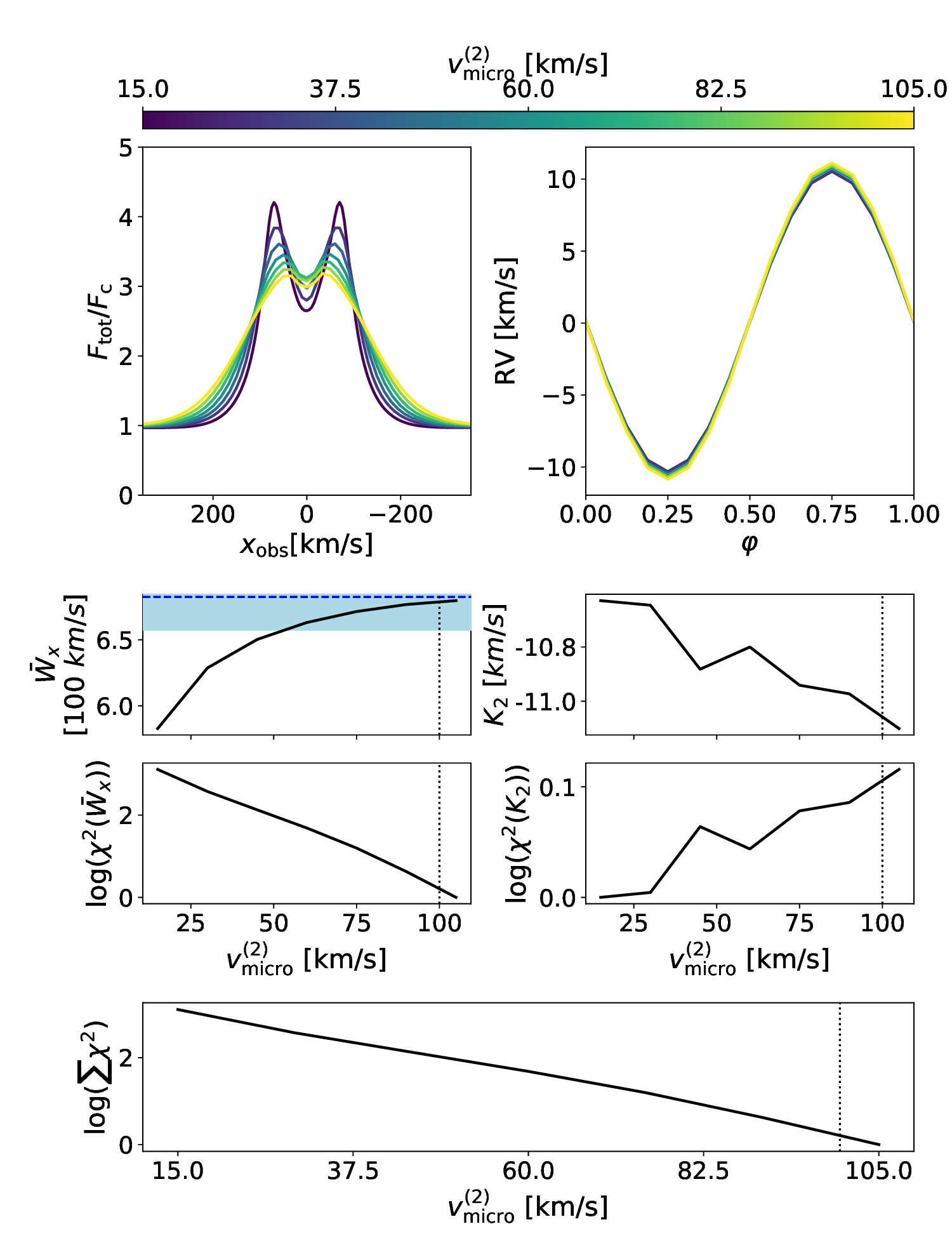}}
  \caption{Parameter study with base parameters defined in
    Table~\ref{tab:params_lb1_fit} for model PB\_SBH. Upper left
    main-panel: Synthetic line profiles at phase $\varphi=0$ (top left
    sub-panel), radial-velocity curve (top right sub-panel),
    equivalent width with corresponding reduced $\chi^2$-value (middle
    left sub-panels), semi-amplitude of the radial velocity curve with
    corresponding reduced $\chi^2$-value (middle right sub-panels),
    and total reduced $\chi^2$-value (bottom panel), all as a function
    of the disc's base density, $\rho_0^{\rm (2)}$. The black dotted
    line indicates our best-fit by eye (see
    Sect.~\ref{subsec:lb1_bh}), and the blue dashed lines display the
    observed values with corresponding error-bars (blue-shaded
    area). Similarly, the top right to bottom right main-panels
    indicate the dependence of our measurable quantities on the disc's
    temperature, $T_{\rm disc}^{(2)}$, the slope of the density
    stratification, $\beta_{\rm D}^{(2)}$, and the micro-turbulent
    velocity, $\vmicro^{\rm (2)}$.}
%
\label{fig:parameter_study_pbsbh}
\end{figure*}
Figure~\ref{fig:parameter_study_pbsbh} displays the results of our
parameter study for the B-star and BH-disc scenario, with variations of
the four free parameters, namely the disc's base density, $\rho_0^{\rm
  (2)} \in 3 \cdot [10^{-13}, 10^{-11}] \,\gcmc$, the slope of the
density stratification, $\beta_{\rm D}^{(2)}\in [1.5, 4]$, the disc's
temperature $T_{\rm disc}^{\rm (2)}\in[6,15]\,\kK$, and the
micro-turbulent velocity, $\vmicro^{\rm (2)}\in
[15,105]\,\kms$. Additionally, we display the reduced $\chi^2$-values
of the equivalent width and the radial-velocity semi-amplitude as
compared to the observations, as well as the combined reduced
$\chi^2$-value (with an equal weight to both quantities).
\paragraph{Impact of $\rho_0^{\rm (2)}.$}
The strength of the line (and thus also the equivalent width)
increases with increasing base density of the disc. Indeed, the
equivalent width follows roughly a quadratic dependence, following the
$\rho^2$-dependence of the opacity for recombination lines. For low
densities, only the photospheric line profile from the primary object
(the B star in this scenario) is visible. Thus, the radial velocity
curve when measured from the \Ha-line wings follows the B-star orbit
($K_1 = 53\,\kms$) for low base densities, and the BH orbit ($K_2 =
7.4\,\kms$) for high base densities, as expected. Additionally, the
overestimation of the radial-velocity amplitude due to the anti-phase
motion of the stellar absorption profile (see
\citealt{AbdulMasih2020}) is clearly observed within our models at
intermediate densities.
\paragraph{Impact of $T_{\rm disc}^{\rm (2)}.$}
The strength of the \Ha~line decreases with increasing disc
temperature. This effect can be explained as follows. With increasing
temperature, the occupation numbers of the lower level become
diminished due to the Boltzmann factor. Thus, the opacity is
considerably reduced, and the disc essentially becomes transparent in
\Ha~at high temperatures. We emphasize that also at temperatures
$T_{\rm disc}^{\rm (2)} \lesssim 6 \,\kK$ (not shown here), the
opacity drops significantly since most hydrogen will be found in the
ground state.  As before, the radial velocity curve measured from the
\Ha~line follows the BH orbit for low temperatures (and thus a strong
disc emission). With increasing disc temperature, the semi-amplitude
of the radial velocity curve becomes overestimated when measured from
the \Ha-line wings due to the anti-phase motion of the stellar
absorption profile. In contrast to the low-density regime described
above, the \Ha-line wings are mainly controlled by the disc emission
at the upper temperature limit, and the radial-velocity curve is
therefore following the BH orbit for the full parameter space
considered here.
\paragraph{Impact of $\beta_{\rm D}^{\rm (2)}.$}
With increasing slope of the disc's density stratification, the
\Ha-line profile switches rapidly from showing a significant disc
emission to the (almost) pure absorption profile from the B
star. Clearly, for $\beta_{\rm D}^{\rm (2)}\gtrsim 2$, the disc
emission vanishes due to decreased densities (in particular at large
distances from the BH and therefore low velocities). The corresponding
radial velocity curves are slightly underestimating the true orbital
velocity of the B star ($K_1=53\,\kms$) due to a slight emission from
the disc in the \Ha-line wings (probing the high-velocity regions and
thus the inner parts of the disc).
\paragraph{Impact of $\vmicro^{\rm (2)}$.}
Both the strength of the line and the radial velocity curve are only
mildly affected by the micro-turbulent velocity in the disc. The shape
of the \Ha-line profile, however, switches from a relatively narrow
and strongly pronounced double-peaked feature at low $\vmicro^{\rm
  (2)}$ to a broad and smooth one at high $\vmicro^{\rm (2)}$. When
compared to observations (see Sect.~\ref{sec:lb1}), the latter is to
be preferred.
\paragraph{Investigation of $\chi^2$.}
The $\chi^2$-values shown in Fig.~\ref{fig:parameter_study_pbsbh}
indicate that both the slope of the density stratification and the
micro-turbulent velocity are relatively well constrained to the
best-fit by eye values found in Sect.~\ref{subsec:lb1_bh}. There
exists, however, a slight degeneracy in the disc's base density and
temperature. For instance, one might find a similar good model by
decreasing both the disc density and the disc temperature at the same
time. With such a model, however, the observed anti-phase behaviour of
line core and line wings (see Fig.~\ref{fig:profiles_best}) could not
be reproduced either. Moreover, at temperatures $T_{\rm
  disc}^{(2)}\lesssim \, 6 \kK$ (not shown here) and $T_{\rm
  disc}^{(2)}\gtrsim \, 15 \kK$, the disc's base density would need to
be increased to obtain a significant equivalent width, translating
then to an X-ray bright disc that has not been observed.
\subsection{Parameter study for the stripped-B and Be scenario with two discs}
\label{app:parameter_study_pstrsbe02}
\begin{figure*}[t]
  \centering
  \resizebox{0.33\hsize}{!}{\includegraphics{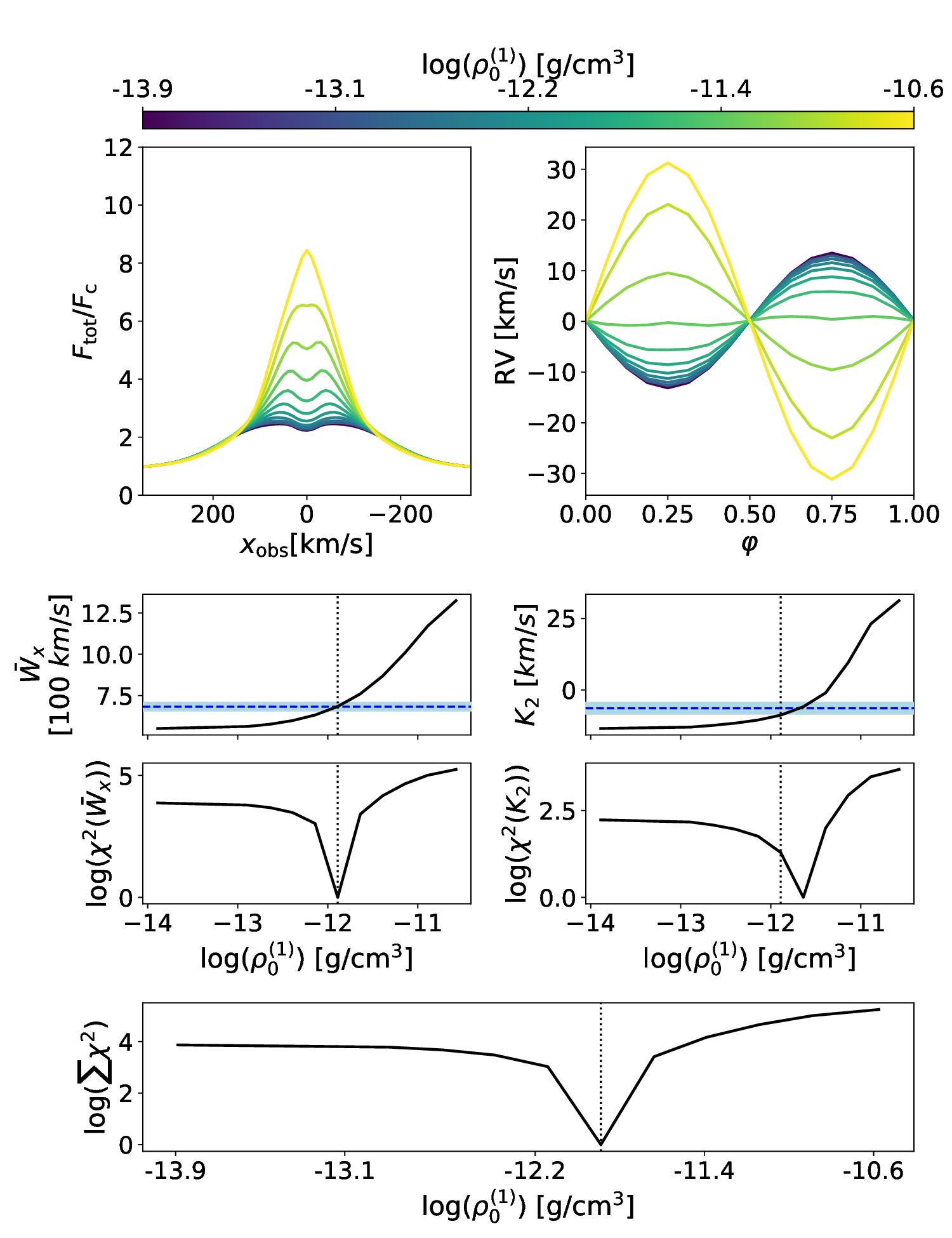}}
  \resizebox{0.33\hsize}{!}{\includegraphics{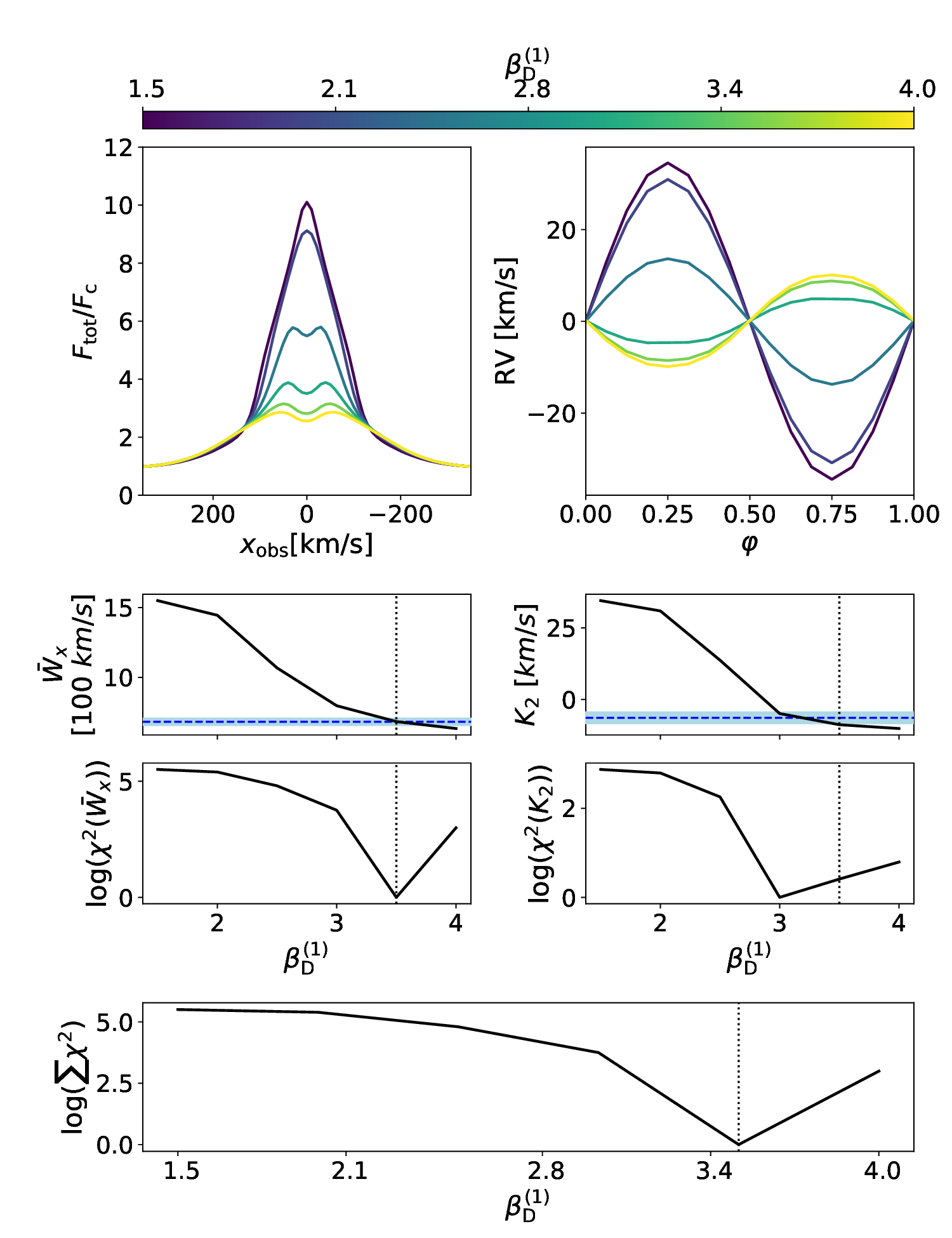}}
  \resizebox{0.33\hsize}{!}{\includegraphics{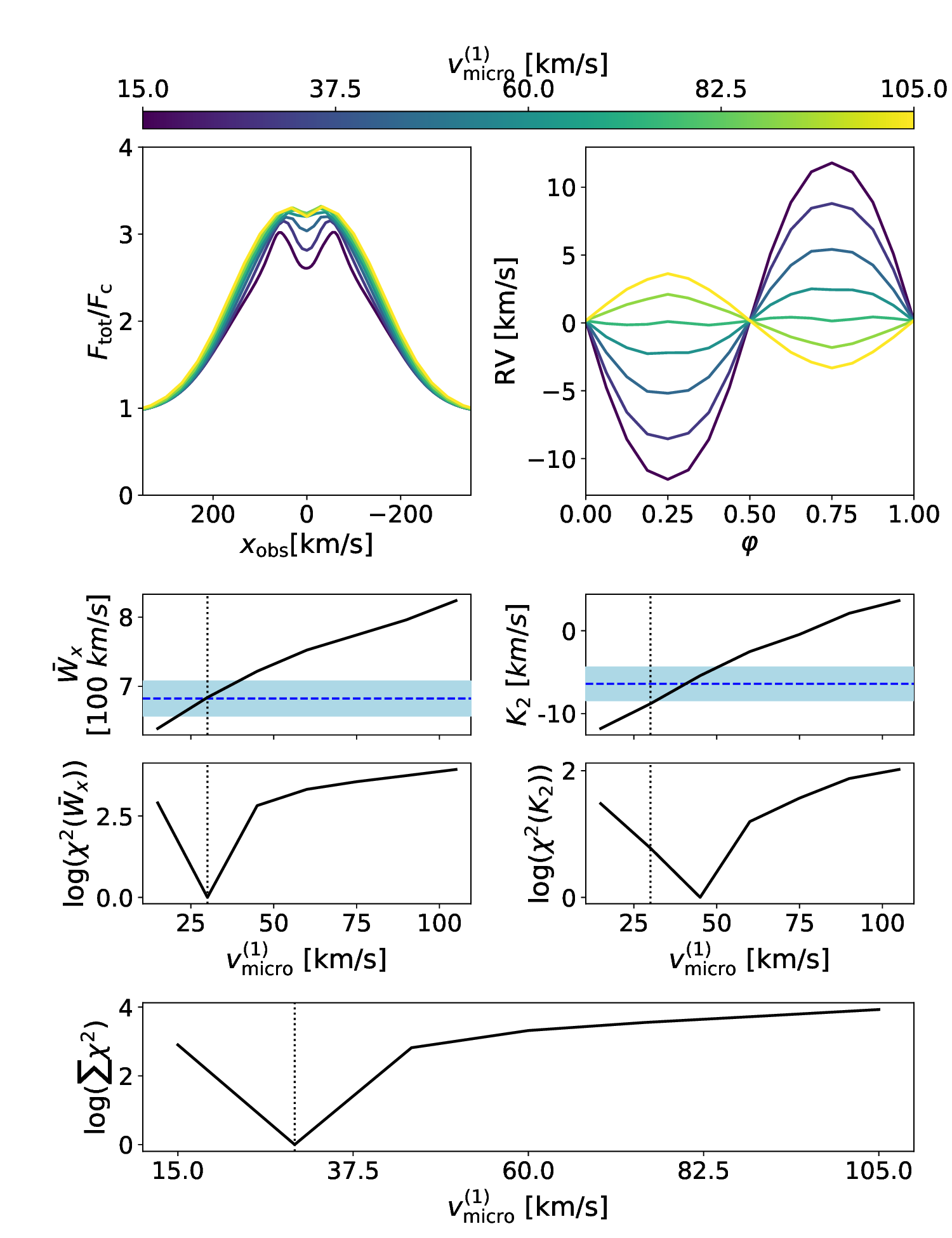}}
  \resizebox{0.33\hsize}{!}{\includegraphics{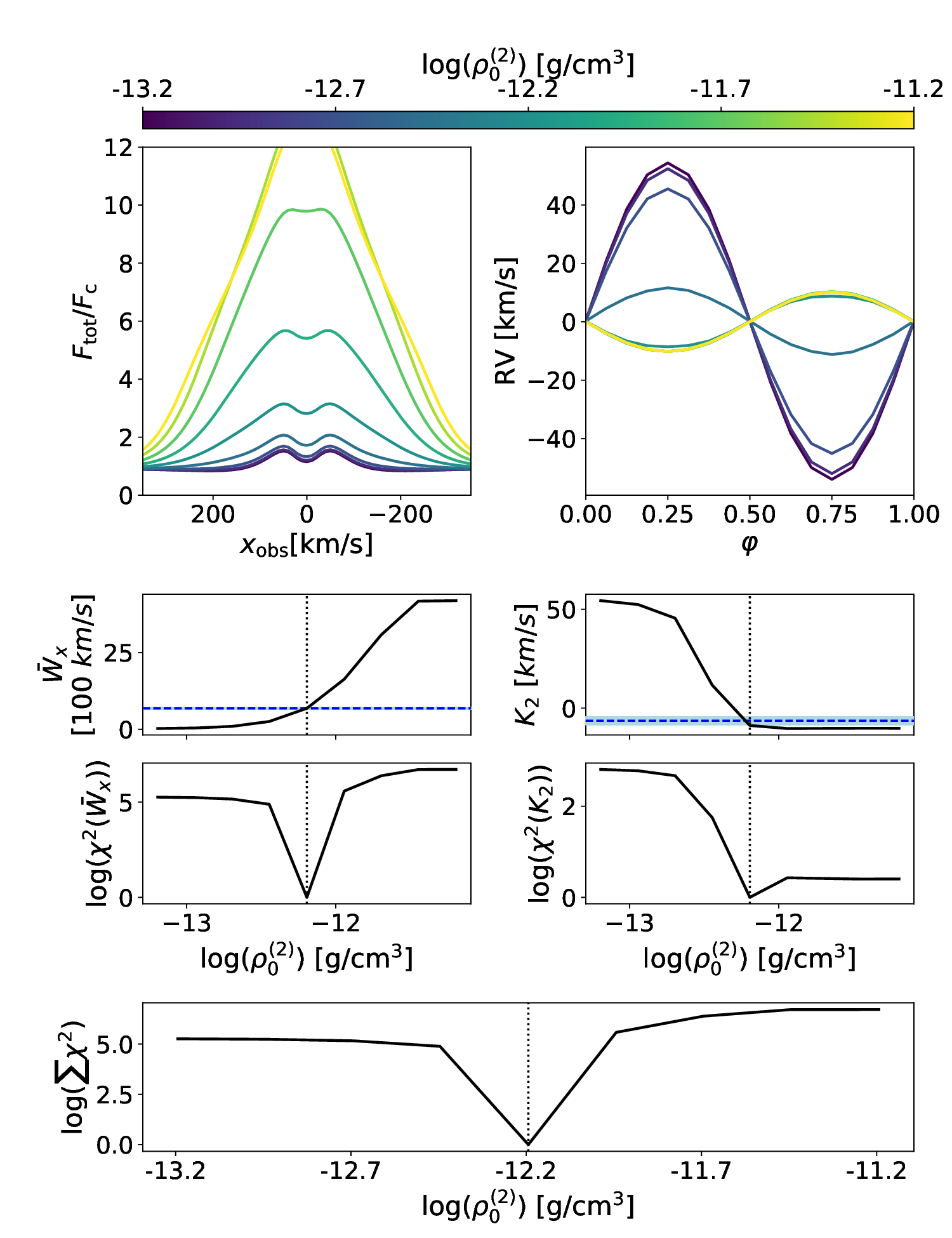}}  
  \resizebox{0.33\hsize}{!}{\includegraphics{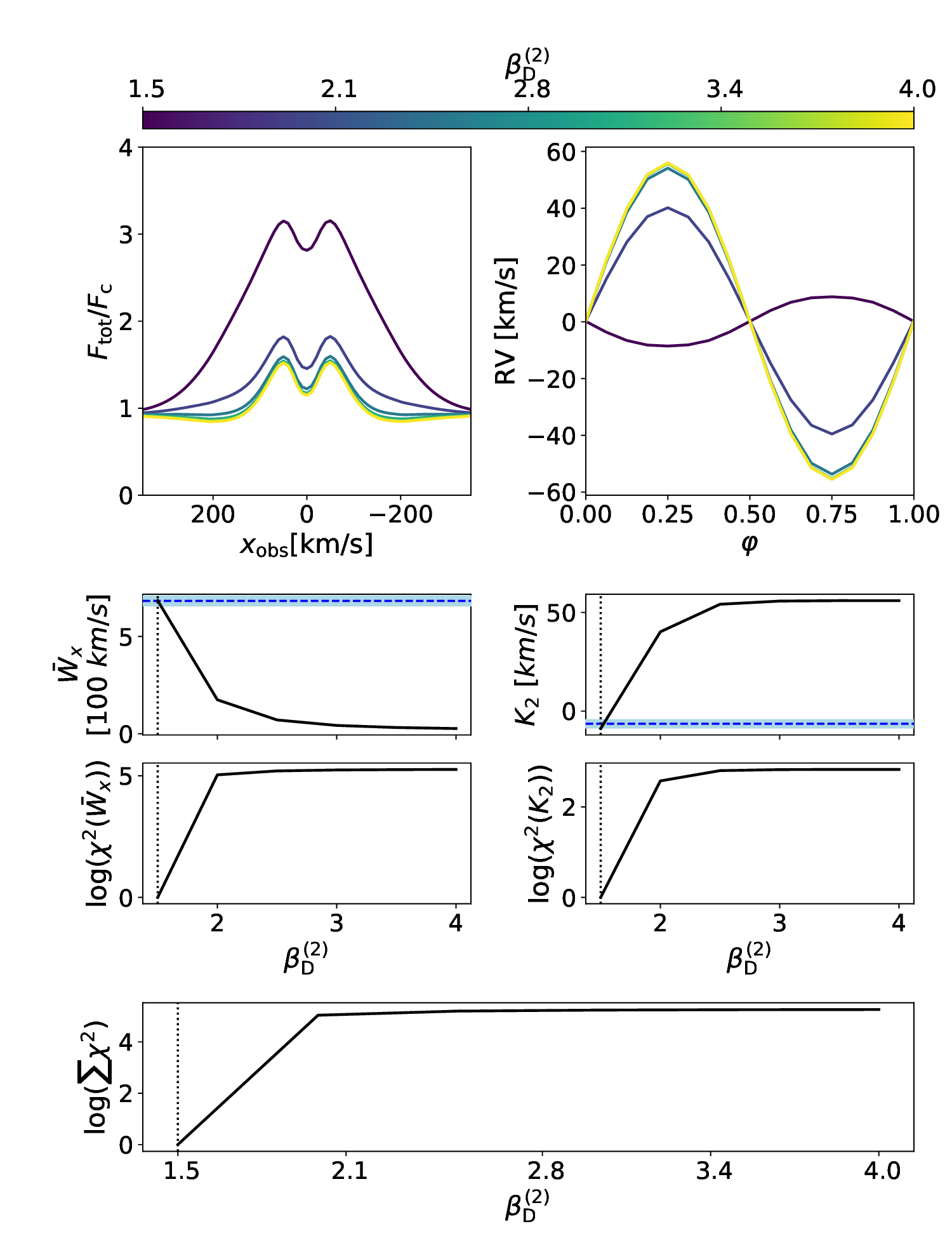}}  
  \resizebox{0.33\hsize}{!}{\includegraphics{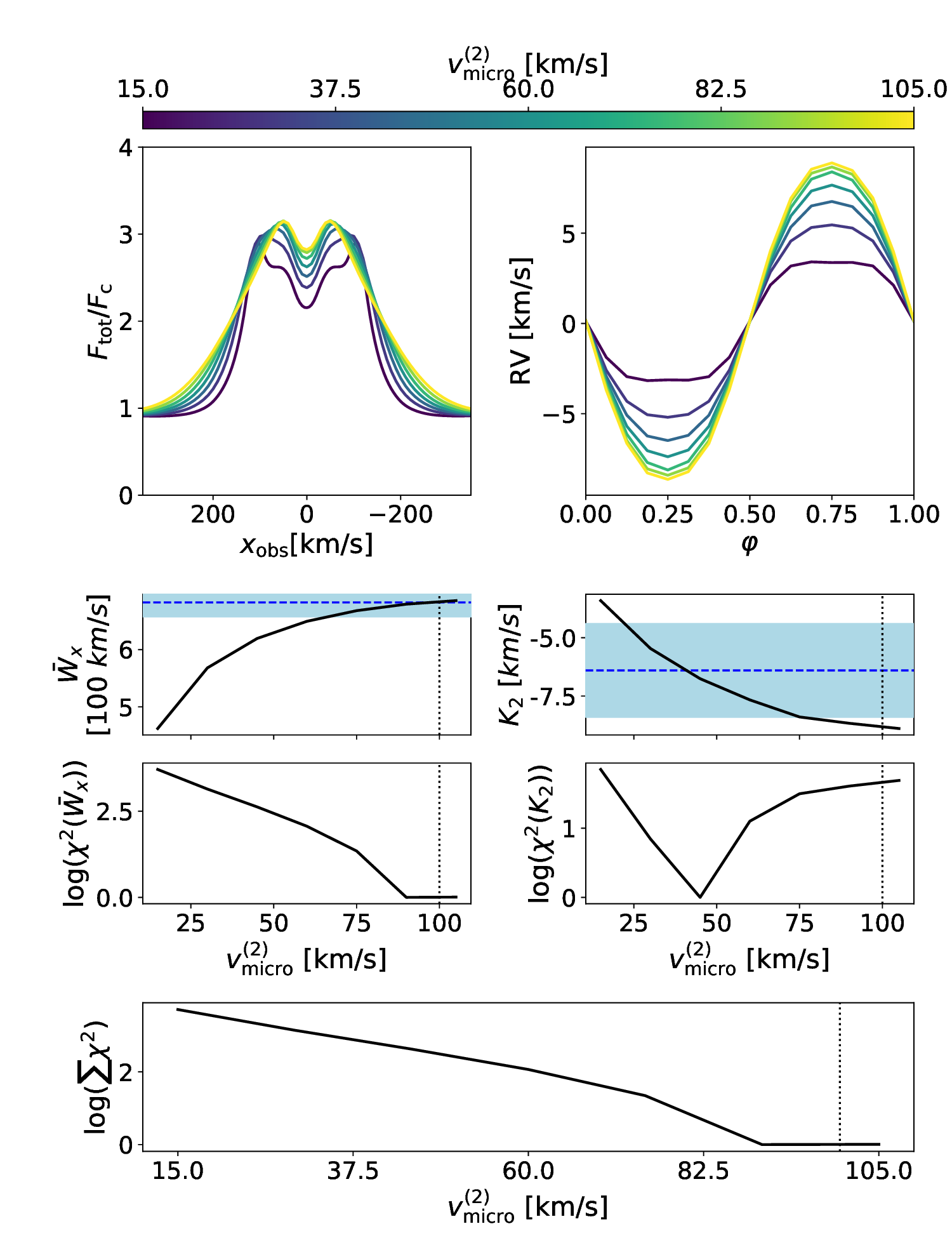}}  
  \caption{As Fig.~\ref{fig:parameter_study_pbsbh}, however for model
    PST\_SBE02 and focussing on the six free parameters $\rho_0^{\rm
      (1)}$, $\rho_0^{\rm (2)}$, $\beta_{\rm D}^{(1)}$, $\beta_{\rm D}^{(2)}$, $\vmicro^{(1)}$, and $\vmicro^{(2)}$.}
%
\label{fig:parameter_study_pstrsbe02}
\end{figure*}
Figure~\ref{fig:parameter_study_pstrsbe02} displays the results of our
parameter study for the stripped-B and Be scenario with two discs, where
the discs are described by six free parameters in total, namely the
base density for each disc, $\rho_0^{\rm (1)} \in [10^{-14},
  10^{-11}]\,\gcmc$ and $\rho_0^{\rm (2)}\in 6 \cdot
[10^{-13},10^{-11}]\,\gcmc$, the slope of the density stratification,
$\beta_{\rm D}^{(1)}$ and $\beta_{\rm D}^{(2)}$ (both in a range $\in
[1.5,4]$), and the micro-turbulent velocities, $\vmicro^{(1)}$,
$\vmicro^{(2)}$ (both in a range $\in [15,105]\,\kms$). As before, we
additionally display the reduced $\chi^2$-values calculated for the
equivalent width, the radial velocity semi-amplitude, and the equally
weighted sum of both.
\paragraph{Impact of $\rho_0^{\rm (1)}$ and $\rho_0^{\rm (2)}$.}
The strength of the line increases with increasing base density of the
stripped-star disc, $\rho_0^{\rm (1)}$, particularly in the
low-velocity regions (due to the overall low velocities in the
stripped-star disc).  At the lower end of our parameter space, the
stripped-star disc becomes transparent and the \Ha-line profile
consists only of three components, namely the absorption profiles of
both stars in the binary system, and the disc emission from the
secondary object (the Be-star disc). Consequently, the radial-velocity
curves vary significantly. For low $\rho_0^{\rm (1)}$, the \Ha-line
wings are probing the Be-star orbit (including the effects of all
underlying stellar absorption profiles, of course). With increasing
$\rho_0^{\rm (1)}$, the radial-velocity amplitude becomes reduced and
switches sign at the upper limit of our parameter space due to a
significant contamination of the \Ha-line emission from the
stripped-star disc.

In contrast to the effects of $\rho_0^{\rm (1)}$, the strength of the
\Ha-line profile increases in both the low and high velocity regions
with increasing $\rho_0^{\rm (2)}$. At the lower end of our parameter
space, the line profiles consist only of the emission from the
stripped-star disc (and again the stellar absorption
components). Thus, the radial velocity curves as obtained from the
\Ha-line wings follow the stripped-star orbit at low $\rho_0^{\rm
  (2)}$, while the anti-phase behaviour probing the Be-star orbit is
measured at high $\rho_0^{\rm (2)}$.
\paragraph{Impact of $\beta_{\rm D}^{\rm (1)}$ and $\beta_{\rm D}^{\rm (2)}$.}
The strength of the \Ha-line profile increases with decreasing slope
of the stripped-star disc's density stratification, $\beta_{\rm
  D}^{\rm (1)}$, in particular in the low-velocity regions. As in
Sect.~\ref{app:parameter_study_pbsbh}, increasing $\beta_{\rm D}^{\rm
  (1)}$ translates to decreasing the density of the stripped-star
disc, particularly in the outer (low-velocity) regions. Thus, the
synthetic line profiles become slightly broader at the upper limit of
$\beta_{\rm D}^{\rm (1)}$. Since the stripped-star disc essentially
becomes transparent at high $\beta_{\rm D}^{\rm (1)}$, the obtained
radial-velocity curve follows the Be-star orbit. Accordingly, the
stripped-star orbit is being probed at low $\beta_{\rm D}^{\rm (1)}$,
with a significant contamination from the Be-star disc though.

Also with decreasing slope of the Be-star disc's density
stratification, the strength of the \Ha-line profile increases, now
within both the low and high velocity regions though. At the upper
limit of $\beta_{\rm D}^{\rm (2)}$, the Be-star disc becomes
transparent and only the stripped-star disc is visible (together with
the stellar absorption profiles, again). Thus, a high $\beta_{\rm
  D}^{\rm (2)}$ is favourable for measuring the stripped-star orbit,
while a low $\beta_{\rm D}^{\rm (2)}$ would give the Be-star orbit
(still contaminated by the stellar absorption profiles).
\paragraph{Impact of $\vmicro^{\rm (1)}$ and $\vmicro^{\rm (2)}$.}
While the equivalent width of the \Ha-line profile is only slightly
increasing with increasing micro-turbulent velocity of the
stripped-star disc ($\vmicro^{\rm (1)}$), the shape of the line
changes significantly from a double-peaked feature to a relatively
smooth and slightly broader appearance. Thus, the radial-velocity
curve is probing primarily the Be-star disc at the lower limit of
$\vmicro^{\rm (1)}$, and the semi-amplitude is increasing with
$\vmicro^{\rm (1)}$ since the line-profile wings become increasingly
contaminated from the stripped-star disc emission.

With increasing micro-turbulent velocity of the Be-star disc
($\vmicro^{\rm (2)}$) now, the equivalent width of the \Ha-line
profile is again only slightly increasing while the shape of the line
changes significantly from a relatively narrow double-peaked line
profile to a much smoother and broader one. Thus, the radial-velocity
curves are probing primarily the Be-star disc at the upper limit of
$\vmicro^{\rm (2)}$. With decreasing micro-turbulent velocity, the
line wings are contaminated by the stripped-star disc, and the
semi-amplitude of the radial-velocity curve therefore becomes
decreased (towards positive values).
\paragraph{Investigation of $\chi^2$.}
As shown by the reduced $\chi^2$-values of our stripped-star and Be-star
models (Fig~\ref{fig:parameter_study_pstrsbe02}), all fit parameters
are fairly well constrained to the best-fit by eye values found in
Sect.~\ref{subsec:lb1a}. There might be a slight degeneracy of the
base density and the slope of the density stratification within each
of the discs, though. For instance, reducing $\rho_0^{\rm (1)}$ and
$\beta_{\rm D}^{(1)}$ (or $\rho_0^{\rm (2)}$ and $\beta_{\rm
  D}^{(2)}$) at the same time might give similar results as found for
our best-fit model.
\end{document}

%% file: definitions.tex
\newcommand{\yhe}{\ensuremath{Y_{\rm He}}}

\newcommand{\msun}{\ensuremath{M_{\odot}}}

\newcommand{\rsun}{\ensuremath{R_{\odot}}}

\newcommand{\Teff}{\ensuremath{T_{\rm eff}}}
\newcommand{\vinf}{\ensuremath{v_{\infty}}}
\newcommand{\mdot}{\ensuremath{\dot{M}}}

\newcommand{\msunyr}{\ensuremath{M_{\odot} {\rm yr}^{-1}}}

\newcommand{\beq}{\begin{equation}}
\newcommand{\eeq}{\end{equation}}
\newcommand{\beqa}{\begin{eqnarray}}
\newcommand{\eeqa}{\end{eqnarray}}

\newcommand{\nbeq}{\begin{equation*}}
\newcommand{\neeq}{\end{equation*}}

\newcommand{\eu}{\ensuremath{{e}}}
\newcommand{\echarge}{\ensuremath{{\rm e}}}

\newcommand{\kms}{\ensuremath{{\rm km}\,{\rm s}^{-1}}}
\newcommand{\gcmc}{\ensuremath{{\rm g}\,{\rm cm}^{-3}}}

\newcommand{\dd}{{\rm d}}

\newcommand{\Ha} {H$_{\rm \alpha}$}

\newcommand{\Rstar}{\ensuremath{R_{\ast}}}

\newcommand{\Mstar}{\ensuremath{M_{\ast}}}
\newcommand{\Lstar}{\ensuremath{L_{\ast}}}

\newcommand{\logg}{\ensuremath{\log g}}

\newcommand{\vth}{\ensuremath{v_{\rm th}}}

\newcommand{\vrot}{\ensuremath{v_{\rm rot}}}

\newcommand{\vsini}{\ensuremath{v{\thinspace}\sin{\thinspace}i}}

\newcommand{\vmicro}{\ensuremath{v_{\rm micro}}}

\newcommand{\vsound}{\ensuremath{v_{\rm s}}}

\newcommand\ie{\hbox{i.e.}}
\newcommand\eg{\hbox{e.g.}}

\newcommand{\vecown}[1]{\ensuremath{\boldsymbol{#1}}}
\newcommand{\matown}[1]{\ensuremath{\boldsymbol{#1}}}

\newcommand{\const}{\ensuremath{\rm const.}}

\newcommand{\vmin}{\ensuremath{v_{\rm min}}}

\newcommand{\indx}[1]{\ensuremath{{#1}}}

\newcommand{\Inu}{\ensuremath{I_{\nu}}}

\newcommand{\ddop}{\ensuremath{\Delta \nu_{\rm D}}}

\newcommand{\kboltz}{\ensuremath{k_{\rm B}}}

\newcommand{\clatitude}{\ensuremath{\Theta}}
\newcommand{\azimuth}{\ensuremath{\Phi}}
\newcommand{\kK}{\ensuremath{{\rm kK}}}
\newcommand{\kelvin}{\ensuremath{{\rm K}}}

\newcommand{\mel}{\ensuremath{m_e}}

\def\apjl{ApJ}%
\def\apjs{ApJS}%
\def\ao{Appl.~Opt.}%
\def\aap{A\&A}%